\documentclass[fleqn,usenatbib]{mnras}

\usepackage[T1]{fontenc}

\DeclareRobustCommand{\VAN}[3]{#2}
\let\VANthebibliography\thebibliography
\def\thebibliography{\DeclareRobustCommand{\VAN}[3]{##3}\VANthebibliography}

\usepackage[normalem]{ulem}
\usepackage{multirow}
\usepackage{graphicx}	
\graphicspath{{../figures/}{../figures_classification/}}
\usepackage{amsmath}	
\usepackage{amssymb}	
\usepackage{diagbox}
\usepackage{xcolor}
\usepackage{newtxtext,newtxmath}

\usepackage{xcolor}

\newcommand{\fermi}{\textit{Fermi}-LAT }

\title{Classification of \textit{Fermi}-LAT sources with deep learning using energy and time spectra}

\author[T. Finke et al.]{
Thorben Finke,$^{1}$ Michael Kr\"amer,$^{1}$ Silvia Manconi,$^{1}$\thanks{Corresponding author, E-mail: manconi@physik.rwth-aachen.de}\\
$^{1}$Institute for Theoretical Particle Physics and Cosmology, RWTH Aachen University, D-52056 Aachen, Germany
\vspace{0.2cm}
\\
Preprint number: TTK-20-45
}

\date{Accepted 17 August 2021. Received 17 August 2021; in original form  14 December 2020}

\pubyear{2021}

\begin{document}
\label{firstpage}
\pagerange{\pageref{firstpage}--\pageref{lastpage}}
\maketitle

\begin{abstract}
Despite the growing number of gamma-ray sources detected by the \textit{Fermi}-Large Area Telescope (LAT), about one third of the sources in each survey remains of uncertain type.  
We present a new deep neural network approach for the classification of unidentified or unassociated gamma-ray sources in the last release of the \textit{Fermi}-LAT catalogue (4FGL-DR2) obtained with 10 years of data. 
In contrast to previous work, our method directly uses the measurements of the photon energy spectrum and time series as input for the classification, instead of specific, human-crafted features. 
Dense neural networks, and for the first time in the context of gamma-ray source classification recurrent neural networks, are studied in depth.
We focus on the separation between extragalactic sources, i.e.\ Active Galactic Nuclei, and Galactic pulsars, and on the further classification of pulsars into young and millisecond pulsars. 
Our neural network architectures provide powerful classifiers, with a performance that is comparable to previous analyses based on human-crafted features.
Our benchmark neural network predicts that of the sources of uncertain type in the 4FGL-DR2 catalogue, 1050 are Active Galactic Nuclei and 78 are Galactic pulsars, with both classes following the expected sky distribution and the clustering in the variability-curvature plane. 
We investigate the problem of sample selection bias by testing our architectures against a cross-match test data set using an older catalogue, and propose a feature selection algorithm using autoencoders. 
Our list of high-confidence candidate sources labelled by the neural networks provides a set of targets for further multiwavelength observations addressed to identify their nature. 
The deep neural network architectures we develop can be easily extended to include specific features, as well as multiwavelength data on the source photon energy and time spectra coming from different instruments.
\end{abstract}




\section{Introduction} \label{sec:intro}

The \textit{Fermi Gamma-ray Space Telescope}, now operating for  more than ten years, has revolutionised our knowledge of non-thermal gamma-rays produced in our Galaxy and beyond.
Its primary instrument, the \textit{Fermi} Large Area Telescope (LAT), is a pair-conversion telescope which surveys the entire sky, detecting gamma-rays with energies from few tens of MeV up to the TeV range \citep{2009ApJ...697.1071A,2012ApJS..203....4A}.
The data obtained by \fermi has led  to the discovery of more than five thousand Galactic and extragalactic point-like (and few extended) gamma-ray sources, enlarging the known gamma-ray source populations to new objects such as radio-quiet pulsars, millisecond pulsars, and globular clusters \citep{Fermi-LAT:2019yla,Ballet:2020hze}. 
All-sky gamma-ray data permitted also to characterise Galactic diffuse emissions \citep{2012ApJ...750....3A,2016ApJS..223...26A} and extragalactic diffuse backgrounds with unprecedented accuracy \citep{2015ApJ...799...86A,2015PhR...598....1F}.
 
Observed sources are distinguished according to their position and spectral characteristics, and have been collected in a succession of  public  source catalogues. 
The fourth  \textit{Fermi}-LAT gamma-ray catalogue, Data Release 2 (4FGL-DR2) \citep{Ballet:2020hze} is the latest  version, and it is based on the first ten years of data taking with gamma-ray energies ranging from 50~MeV to 1~TeV. The 4FGL-DR2 catalogue is obtained with the same analysis methods as the 4FGL catalogue \citep{Fermi-LAT:2019yla}, but considering two additional years of data and thus adding 723 new sources, resulting in a total of 5788 detected objects.  

The most numerous and brightest source population is currently represented by Active Galactic Nuclei (AGN), where jets originating from a supermassive black hole at the centre of a galaxy emit electromagnetic radiation in a broad  range, from radio frequencies to gamma-rays with TeV energies \citep{1995PASP..107..803U,Padovani:2017zpf}. More information on the various classes of AGN detected by \fermi is contained in dedicated catalogues \citep{Fermi-LAT:2019pir}. 

Aside from AGN, the second main class of sources contained in present catalogues consists of gamma-ray pulsars (PSR), divided further into young pulsars (YNG) and millisecond pulsars (MSP) \citep{TheFermi-LAT:2013ssa}. 
The gamma-ray emission of a pulsar can be attributed to a rapidly rotating, highly magnetised  neutron star surrounded by a plasma magnetosphere, although the exact emission mechanism and the location of the emission region are still uncertain \citep{1998ApJ...508..328H,Romani159}.  
MSP, or so-called recycled pulsars, are believed to be old pulsars for which the period has decreased via accretion from a companion object \citep{2009Sci...325..848A}. 

The AGN and PSR classes of gamma-ray emitters are characterised by different timing and spectral properties. 
While AGN typically have a variable gamma-ray flaring pattern on time scales of months, PSR are instead non-variable on long time scales \citep{TheFermi-LAT:2013ssa}. 
Moreover, the energy spectra of AGN are generically well described by a power law shape, which is softer than that of a PSR. The energy spectra of PSR have  a spectral curvature, where a break is observed at low and high  energies in the \fermi energy band, requiring the addition of an exponential cutoff at a few~GeV. 

In the 4FGL-DR2 catalogue the gamma-ray sources are identified as, or associated with, an AGN, a PSR, or some more rare population class. Sources are identified according to a correlated timing signature at different wavelengths, or only associated requiring positional coincidence with a counterpart source at other wavelengths. A total of 1794 sources (about one third) in the 4FGL-DR2 catalogue remains unassociated or unidentified (for brevity, we will label both unassociated and unidentified sources as UNC, \textit{unclassified}, in what follows). 
The fraction of UNC sources was similar in previous catalogues \citep{Acero:2015gva}. 
The distribution of their spectral properties and their Galactic latitude reveals the composite nature of the UNC population \citep{Fermi-LAT:2019yla}. 

Given the rapid growth in size and complexity of astronomical datasets, machine learning algorithms have become increasingly powerful tools, in particular for object classification tasks, see e.g.\ \citet{Ball:2009wd,baron2019machine}. The number of gamma-ray sources detected by \textit{Fermi}-LAT has increased from 1451 in the first catalogue  \citep{Abdo_2010} to 5788 in the most recent 4FGL-DR2 catalogue based on ten years of the \fermi mission. Because of the large amount of data, and the subjective nature of classification based on human-crafted features, various authors have developed automated schemes based on machine learning techniques in order to classify  gamma-ray sources.
These approaches are generally based on source features collected in gamma-ray catalogues only, and provide a list of particularly interesting candidate target sources for follow-up observations. Machine learning has been applied to  investigate the nature of the UNC sources in current and past \textit{Fermi}-LAT catalogues \citep{Mirabal:2012em,Mirabal_2016,Saz_Parkinson_2016,Hui:2020cmv,Luo_2020}, or to predict the 
subclass of AGN of uncertain type \citep{Doert_2014,Chiaro:2016noj,Salvetti_2017, Kovacevic:2020sly}. 
Furthermore, machine learning based analyses are instrumental to population studies,  pointing to new source classes \citep{Saz_Parkinson_2016} or to a more exotic origin of gamma-ray emission, such as dark-matter sub-halos \citep{Mirabal:2012em}. 

The studies thus far  typically use standard machine learning algorithms, such as classification trees, logistic regression, or random forest classifiers, and are based on the features provided by the \fermi catalogues, or simple additional  ones, such as hardness ratios, constructed from the energy and time dependent gamma-ray fluxes, see e.g.\ \cite{Saz_Parkinson_2016}. Such an approach requires a careful choice of the features to consider, introducing ambiguities and potential biases e.g.\ due to existing knowledge of the source sample. 
To automatise  and improve the  performance, machine learning techniques for feature selection were recently applied to the classification of 4FGL sources \citep{Luo_2020}. The analysis by \citet{Luo_2020} demonstrates that feature selection by machine learning can improve the classification performance compared to analyses where the choice of features is guided by prior knowledge. The result of this approach depends on the particular release of the catalogue used, as  the number and choice of features may change (e.g.\ the introduction of a new spectral form). 
Moreover, the features might not represent all the information encoded in the spectral data, and the physical interpretation of the importance of some features for the classifier, for example errors in the flux measurement, may be difficult. 
  
In this paper we propose a different approach to classify UNC gamma-ray sources.  We employ a fully connected deep neural network and a recurrent neural network trained directly on the energy and time dependent fluxes presented in the Fermi-LAT catalogue, instead of using human-crafted features as input.

A  related deep learning approach based on energy and time dependent fluxes  has recently been used to predict the subclass of AGN of uncertain type in \cite{Kovacevic:2020sly}. 
Our work goes substantially beyond \cite{Kovacevic:2020sly}: We significantly extend the investigation of deep learning methods and the corresponding numerical analysis, exploring recurrent neural networks and applying deep learning to the categorisation of UNC sources for the first time.

We here concentrate on binary-type classification of UNC sources in two steps by using dense and recurrent deep neural networks. In particular, we first perform interclass binary-type classification, by predicting if UNC are AGN or PSR, and then PSR intraclass binary-type classification separating YNG vs. MSP. 
We demonstrate that deep neural networks based on energy and time spectra provide  powerful classifiers to separate AGN from PSR, as well as YNG from MSP, with a performance that is comparable with previous feature-based analyses presented in the literature.
In addition, our method is very flexible and can easily be generalised  to include multiwavelength data for the energy and time spectra coming from different observatories. 
We also reconsider the classification of gamma-ray sources based on features, and propose a  procedure based on autoencoders.

The paper is organized as follows. 
The dataset is introduced in Sec.~\ref{sec:data}.
Our methods, including the main neural network architectures and the training and testing strategies are presented in Sec.~\ref{sec:method}. In Sec.~\ref{sec:results} we discuss the classification performance on labelled catalogue sources, a cross-matching procedure using an older catalogue, and we provide  our predictions for the unlabelled sources in the current \fermi catalogue. 
In Sec.~\ref{sec:autoenc} we introduce an alternative approach for feature selection using autoencoders, before concluding in Sec.~\ref{sec:theend}. A link to a repository containing ancillary files with our classification results, together with more details on the dataset used are provided in the Data Availability section at the end of the manuscript.

\section{Dataset} \label{sec:data}

The central idea of our classification procedure is to directly use the gamma-ray energy spectra and time series, rather than a selection of derived features. The reference dataset is the 4FGL-DR2 catalogue\footnote{Publicly available at \url{https://fermi.gsfc.nasa.gov/ssc/data/access/lat/10yr_catalog/}, see section Data Availability.}, which has been obtained with the same analysis methods as the 4FGL \fermi catalogue \citep{Fermi-LAT:2019yla}, but considering two additional years of data and thus updating spectral parameters, spectral energy distributions and source associations.

Each source in a gamma-ray catalogue is  described by measurements as well as human-crafted features (a total of 74 in the 4FGL-DR2 catalogue) designed to represent the main source properties.
Apart from the features connected to the source position in the sky, many features describe the energy and timing properties of photon fluxes.  
The 4FGL-DR2 catalogue reports the photon fluxes in seven energy bands, as well as the integrated flux in the full energy range of the analysis.
The photon flux is then fitted with three different spectral shapes over the full energy range, and the best fit parameters, together with their uncertainties, are provided. 
As for the time variability, in the 4FGL-DR2 catalogue one-year light curves (integrated over the full energy range) are reported, along with the variability index, which weights the relative variability between different time bins with respect to the mean value of all bins.

We here use the gamma-ray flux values in the seven energy bands as reported in the 4FGL-DR2 catalogue (entry called \texttt{Flux\_Band} in the catalogue tables, referred to as  energy band/spectra/series data in what follows), together with the corresponding significance in each energy band (\texttt{Sqrt\_TS\_Band}), referred to as 'significance' in what follows.\footnote{The source significance is quantified by the test statistics TS = $2 \log(\mathcal{L}/\mathcal{L}_0)$, where $\mathcal{L}$ and $\mathcal{L}_0$ are the log-likelihood functions with and without the source of interest included in the model, respectively.}
These  values are obtained from the spectral fit over the full energy range, but adjusting the normalisation  in each  band. 

As for the time spectra, we select the ten one-year light curves and the corresponding significances integrated over the full energy range (\texttt{Flux\_History} and \texttt{Sqrt\_TS\_History}, referred to as time spectra/series in what follows). Note that two-month intervals were provided in the first release of the 4FGL catalogue, but removed in the 4FGL-DR2 release, since one-year light curves were demonstrated to capture most of the variability information \citep{Fermi-LAT:2019yla}. 
 
A cross-match of the classification performance based on the 3FGL catalogue \citep{Acero:2015gva} will be presented in Sec.\ref{sec:xval}. Some differences in the structure of the energy and time spectra are present in this case: the energy bands are reduced to five, and the time spectra are represented by monthly bins, instead of yearly bins. 
In particular, the significance for the time spectrum of the gamma-ray flux is not given. In addition, the time spectrum is not given in yearly bins, but in 48 monthly bins. Finally, given the lower statistics, the energy spectrum is only provided for five energy bins. 
So the input dimensions are $(5, 2)$ for the energy spectrum including the significance, and $(48, 1)$ for the timing information (four years with 12 monthly bins each).

While our benchmark architectures use the energy and time spectra, and the corresponding significances, 
we also test some variations, where one of the two  input spectra is substituted by standard features in the catalogue.  
This is done considering, for example, 
the feature named \texttt{Variability\_Index}, 
which quantifies the variability of the gamma-ray flux with one single estimator, or the measured source Galactic latitude (\texttt{GLAT}) (see the definition in \citet{Fermi-LAT:2019yla}).


\section{Method}\label{sec:method}
We use deep neural networks, i.e.\ neural networks composed of several layers, for the classification of gamma-ray sources (for an introduction to deep neural networks and other supervised learning techniques in physical sciences see e.g.\ \citet{Carleo_2019}).
We employ two deep neural network architectures: a fully connected network, also called dense neural network (DNN),   
and a recurrent neural network (RNN). RNNs are designed for analysing sequential data or time-series data, and can process variable length sequences of inputs. They have been used successfully for classification tasks for example in astronomy \citep{2019NatAs...3..680I,2020MNRAS.491.4277M,2018NatAs...2..151N,Hinners_2018,2020MNRAS.493.2981B} and in high-energy physics \citep{Guest_2016,egan2017long,Louppe_2019,Cheng_2018, Fraser_2018,englert2020sensing}. 
In what follows we describe the neural network architectures in detail, focusing in particular on the similarities and the differences between the two.

\subsection{Dense neural network}

\begin{figure}
\includegraphics[trim = 50 50 50 50, clip, width=.45\textwidth]{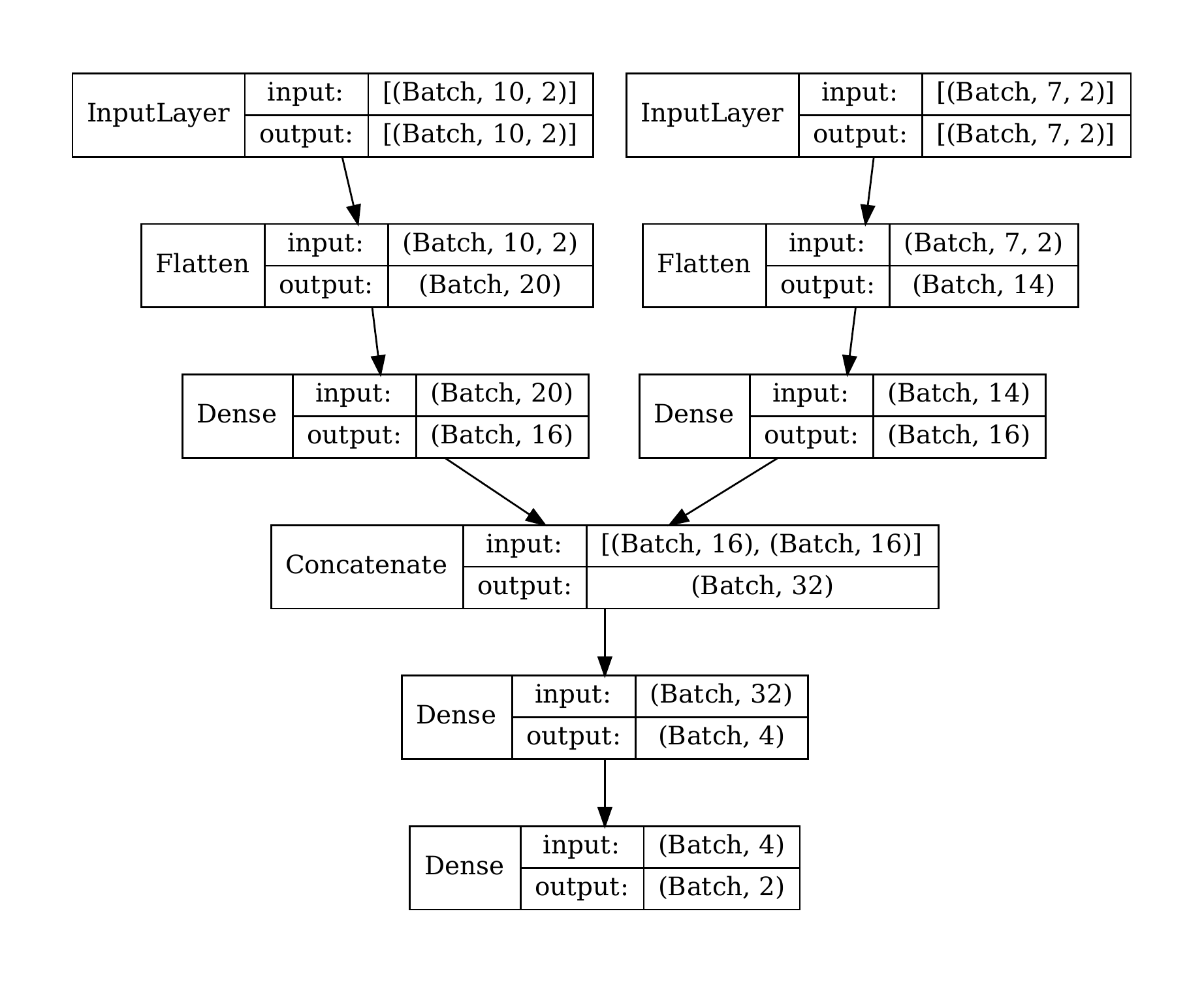}
\caption{Sketch of the dense neural network architecture for the classification of \fermi sources. Each box illustrates one layer, stating the type of layer on the left and on the right the input shape at the top and the output shape at the bottom. The term \textit{Batch} in these shapes correspond to the batch size, which can vary for different network calls.  \label{fig:ml}}
\end{figure}

In a dense  neural network (DNN)  each neuron in a specific layer is  connected to all neurons in the previous and subsequent layers, see e.g.\ \citet{Carleo_2019}. A visualisation of the DNN architecture used in this work is shown in Fig.~\ref{fig:ml} for illustration. Each box in the figure corresponds to a layer in the DNN, and shows the layer type, the input shape to the layer and the shape of its output. In the dense layers we use $\text{ReLU}(x)=\max(0, x)$ as activation function, except for the final layer where the softmax activation converts our output into a probability distribution.
As illustrated in Fig.~\ref{fig:ml}, the DNN consist of two branches, one for the energy spectrum and one for the time series, which are eventually merged into a fully connected network for classification. 
Both our neural network architectures are very flexible and additional features, apart from the energy and time spectra, can easily be included in the classification algorithms. We will explore the impact of additional features on the classification performance in Sec.~\ref{sec:results}, in particular for the task of differentiating between the YNG and MSP classes of pulsars.
When adding additional features, we append a third branch to the architecture, which is merged for classification with the other two branches in the same way as before. 

We have used a ten-fold cross-validation procedure to optimise the hyperparameters of our networks. Hyperparameters are those parameters that are not optimised during training itself, but need to be set beforehand. In this work we considered the number of layers in each branch, the number of nodes in these layers, the number of layers in the final classification part as well as the number of nodes  within these layers as hyperparameters of the architecture. We tune these hyperparameters using ten-fold cross validation as follows: The training set is divided into ten subsets. The networks with different hyperparameters are trained on nine of these subsets and evaluated on the tenth. This is repeated ten times, always taking another subset for the evaluation. This  procedure allows us to evaluate on the full training set while keeping a test set that has not been used for hyperparameter optimisation.

As our data consists of serial data with two features, the input needs to be flattened to be analysed by the DNN. The hyperparameter search gives good results for the DNN with one hidden layer with 16 nodes for both time and energy series. After concatenating the two branches, 32 features are the input to the classification part of the network. This part consists again of one hidden layer, now with  four nodes, and the final output layer with two nodes (one for each class, see next subsections).

\subsection{Recurrent neural network}
Just like the DNN, also the default RNN architecture consist of two branches, one for the energy and one for the time series. The main difference  to the DNN architecture is that the dense layers are now substituted with recurrent layers.
Thus, the first step of flattening the input is not necessary for the RNN, as it is designed to take serial data with several features as input. We find good classification results for three recurrent layers per branch with eight units each. After concatenating the two branches, 16 features are the input to the classification part of the network. This part is the same for the RNN as for the DNN.
The recurrent layers used for the serial data are \textsc{SimpleRNN} layers from \textsc{Tensorflow/Keras} \citep{tensorflow2015-whitepaper,chollet2015keras} which use a hyperbolic tangent activation function.

To summarize, the key difference between the RNN and the DNN is the way the input is treated. For the RNN the input is sequential, i.e.\ the entries in the time and energy series are analysed step by step, with earlier entries influencing the impact of later ones. The DNN takes the input as a whole, and correlations between earlier and later entries in the time and energy series would have to be learned from the data.

\subsection{Data preprocessing}

\begin{figure}
\centering
\includegraphics[clip, width=.5\textwidth]{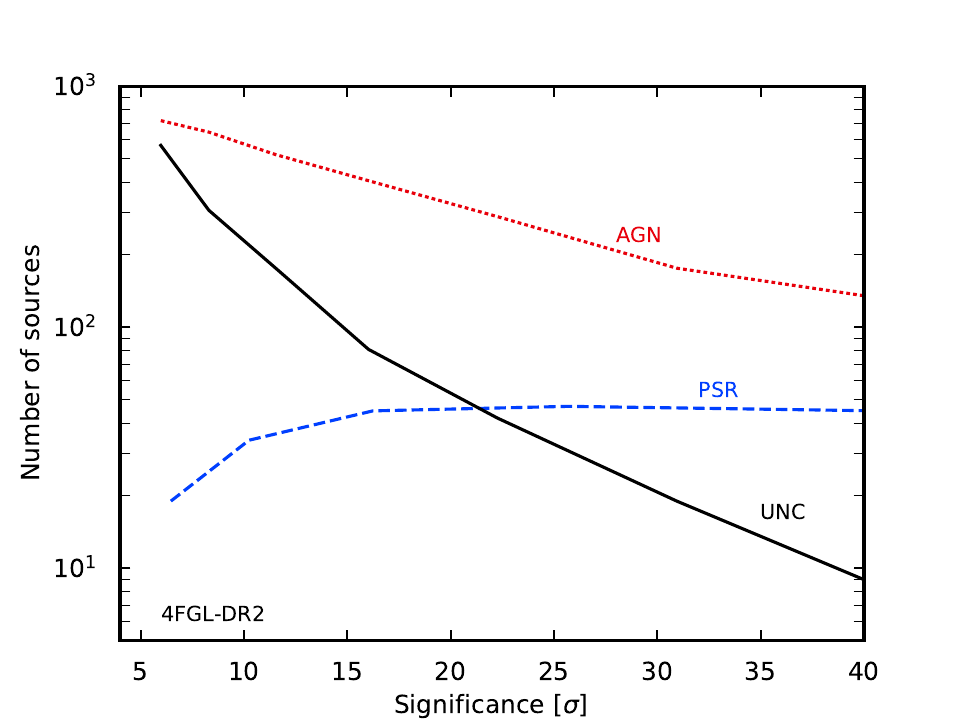}
\caption{Number of sources in the \fermi 4FGL-DR2 catalogue as a function of the source significance (in $\sigma$ units) in the 100~MeV to 1~TeV energy band, for the three main source classes: AGN, PSR and UNC (see text for more details). \label{fig:catstats}}
\end{figure}

As a first step in data preparation, we select sources from the 4FGL-DR2 catalogue falling within the two broad classes of AGN and PSR. 
The PSR class includes all sources that are identified as or associated with pulsars (\texttt{CLASS1=PSR,psr}).\footnote{Source classes are reported in the catalogue in the entry called \texttt{CLASS1} as upper case (i.e. \texttt{PSR}) when they refer to identified sources, while the ones reported as lower case (i.e. \texttt{psr}) refer to associated sources. For the criteria of source association/identification we refer the reader to the Introduction, Sec.~\ref{sec:intro}, and to \citep{Fermi-LAT:2019yla,Ballet:2020hze}.} 
As for the AGN class, we include all sources that are identified as or associated with any type of AGN, including subclasses of blazars as the Flat Spectrum Radio Quasars (FSRQ), BL Lacs (BLL), blazars of unknown type (BCU), and other more rare AGN types, i.e.\ with \texttt{CLASS1=FSRQ, fsrq, BLL, bll, BCU, bcu, CSS, css, RDG, rdg, NLSY1, nlsy1, agn, ssrq, sey}, see \cite{Fermi-LAT:2019yla,Ballet:2020hze} for more details on the AGN subclass definition. We will refer to identified/associated AGN and PSR as labelled sources. The UNC sources are instead those without any identification or association, reported as a blank space in the 4FGL-DR2 catalogue. 
The numbers of the corresponding sources in the 3FGL and the 4FGL-DR2 catalogues are reported in Table~\ref{tab::catalog}. We note an increase of almost a factor of two in the number of sources between the two catalogues. In Fig.~\ref{fig:catstats} we display the number of sources in the 4FGL-DR2 catalogue for each class as a function of the significance (in standard deviations $\sigma$).  The number of UNC sources is subdominant at high significances, but increases monotonically towards lower significances, and is similar to the number of AGN detected at $5\sigma$.

\begin{table}
\caption{Composition of the 4FGL-DR2 and 3FGL catalogues. The number of sources in each source class (see text for definition of AGN, PSR and UNC classes) is reported, together with the total number of sources in each catalogue.}
\centering
\begin{tabular}{ c @{\hspace{10px}} c @{\hspace{10px}} c } \hline\hline

\textbf{Source class }	& \textbf{4FGL-DR2}	& \textbf{3FGL}		\\ \hline
AGN						&  3508				&  1745				\\
PSR						&  259				&  167				\\
UNC						&  1679				&  1010				\\ \hline
Tot						&  5788				&  3034				\\ \hline\hline

\end{tabular}
\label{tab::catalog}
\end{table}


Since we are also interested in the classification of pulsars as YNG or MSP, we use the public list of LAT-detected gamma-ray pulsars\footnote{\url{https://confluence.slac.stanford.edu/display/GLAMCOG/Public+List+of+LAT-Detected+Gamma-Ray+Pulsars}} to build a cross-match between this list and the 4FGL-DR2 catalogue. We obtain a list of 230 known gamma-ray pulsars, split into 123  YNG and 107 MSP. 
Note that energy and time spectra are given for each source, and thus there is no need to filter out sources with missing information, as necessary when using features. The numbers in Table~\ref{tab::catalog} thus represent the full statistics of the source sample used for the classification.

Note that the input to the neural networks consists of the energy spectra and the time series, together with the corresponding significances. For the RNN this amounts to two series with two entries each, i.e.\ the flux and the significance for each bin. In the case of the DNN we reshape the input into two one-dimensional vectors consisting of $2 \times 7$ entries for the energy spectrum and $2 \times 10$ entries for the time series. 
We shift and re-scale each entry such that the mean over the training set is zero and the standard deviation is one.

\subsection{Neural network classification training and testing}\label{subsec:training}

We perform a  two-step binary classification to first separate between the two main source classes in the 4FGL-DR2 catalogue, i.e.\ AGN and PSR, and then between the two types of PSR, i.e. MSP and YNG, as done e.g. by \citep{Saz_Parkinson_2016}.

The training of our neural networks is characterised by the choice of the optimiser, the loss function, the initial learning rate and the number of training updates in terms of batch size and number of epochs. Here, the batch size corresponds to the number of sources passed through the network during one training step, and one epoch of training corresponds to having used every training source once.
We use the Adam optimiser~\citep{adam} to minimise the categorical cross-entropy loss function
\begin{equation}
\text{CE}(y^{\rm{true}}, y^{\rm{pred}}) = \frac{1}{N} \sum_{n=1}^N  
y^{\rm{true}}_{n} \log(y^{\rm{pred}}_{n})+
(1-y^{\rm{true}}_{n}) \log(1-y^{\rm{pred}}_{n})\,.
\label{Eq:crossentropy}
\end{equation}
Here, the sum over $n$ corresponds to a sum over sources and $\log$ refers to the natural logarithm.
The $y^{\rm{true}}_{n}$ is one if source $n$ belongs to a given category, and zero otherwise. The network's prediction of source $n$ to belong to the given category is denoted as $y^{\rm{pred}}_{n}$. 
Since we are looking at (subsequent) binary classifications, the use of categorical cross entropy is equivalent to binary cross entropy.

We find that efficient training is achieved for the DNN with a batch size of 64 and an initial learning rate of 0.01. For the RNN a batch size of 32 and an initial learning rate of 0.001 performs best. We train the networks for 50 and 150 epochs for the task of AGN vs.\ PSR and MSP vs.\ YNG classification, respectively, see also Sec.~\ref{sec:mspyng}.
We train our networks on 70\% of the labelled sources and evaluate on the remaining 30\%. The test split is performed such that the relative amount of the two classes is the same in the training and test sets. The  performance measures and uncertainties stated in this paper correspond to the mean values and the standard deviations of ten training runs. For each run a new network is initialised  and the training and test split is performed randomly. The uncertainties thus combine  those from different network initialisations and from the selection of the training data.

\subsection{Validation of the performance on labelled sources}\label{subsec:measures}

Due to the large imbalance of the two classes in our dataset (see the AGN and PSR numbers in Table~\ref{tab::catalog}), the measures used to quantify the performance of the classifier need to be selected with some care.  
Indeed, the global accuracy  (defined as the ratio of correctly classified sources over the total number of sources) is sensitive to this imbalance. 
For example, for our specific dataset the accuracy would approach already $\approx 93$\% if the classifier simply labelled all the sources as AGN.

In what follows, we call PSR the positive class and AGN the negative one. With this definition, correctly classified PSR are true positives (TP), while false positives (FP) are AGN incorrectly classified as PSR. 
Correctly classified AGN are true negatives (TN), while PSR incorrectly classified as AGN are false negatives (FN). For binary classification, these numbers can be summarised in a so-called confusion matrix. The confusion matrix shows in its rows the numbers of instances with a certain predicted label and in the columns instances with a certain true label. It can thus be written as:

\begin{center}
\begin{tabular}{c|c|c}
					& True AGN	& True PSR	\\ \hline
	Predicted AGN	& TN 		& FN		\\ \hline
	Predicted PSR	& FP 		& TP
\end{tabular}
\end{center}

Additionally, we consider the following performance measures:
\begin{enumerate}
\item Accuracy = $\frac{\displaystyle {\rm TP}+{\rm TN}}{\displaystyle {\rm TP}+{\rm FP}+{\rm TN}+{\rm FN}}$
\item F1 score = $\frac{\displaystyle {\rm TP}}{\displaystyle {\rm TP}+0.5({\rm FP}+{\rm FN})}$
\item Equal accuracy 
\item Cross-entropy loss (see Eq.~\ref{Eq:crossentropy})
\item Area under the Receiver Operating Characteristic (ROC) curve.
\end{enumerate}

We choose PSR as the positive class since we want to further differentiate between the pulsar subclasses MSP and YNG. For the MSP vs.\ YNG classification we arbitrarily refer to MSP as the positive class.

The confusion matrix and threshold dependent metrics are calculated at the decision boundary that gives the highest accuracy on the training set. The decision boundary refers to the threshold in the positive class score necessary for a source to be labelled as belonging to this class.
The F1 score is the harmonic mean of precision (the ratio of correct  to all PSR classifications, i.e.\ TP/(TP+FP)) and recall (the ratio of correct PSR classifications to all actual PSR, i.e.\ TP/(TP+FN)) at the same decision boundary. 
The ``equal accuracy" measure is given by the accuracy of the classifier when operating at the decision boundary that gives the most similar accuracy in both classes on the training set. 
By scanning over all possible decision boundaries of the classifier, one can construct the ROC curve. Here, the ROC curve corresponds to the PSR efficiency (true positive rate) as a function of the AGN efficiency (false positive rate). We recall that the area under the ROC curve (AUC) is bound to be between zero and one. A perfect classifier would have an AUC value of one, while a random classifier would correspond to an AUC value of 0.5. 
AUC values below 0.5 indicate that the classifier is switching the classes. An AUC value of zero corresponds to a classifier that inverts all labels.


\section{Results}\label{sec:results}
In this section we discuss the results obtained on the dataset introduced in Sec.~\ref{sec:data} using the deep network architectures and methods introduced in Sec.~\ref{sec:method}. We first discuss the classification performance on labelled sources in the 4FGL-DR2 catalogue. Then, the performance of our classifiers is tested on a cross-match set of sources between the 3FGL and the 4FGL-DR2 catalogues. Finally, the predictions for UNC sources in the 4FGL-DR2 catalogue are presented, and the main characteristics of the candidate sources found by our deep networks are investigated.

\subsection{Performance on labelled sources in the 4FGL-DR2 catalogue}

\subsubsection{AGN vs.\ PSR classification}\label{sec:agnpsr}

We first discuss the AGN vs.\ PSR classification. Table~\ref{tab::results4FGL} summarises the results on the labelled data of the 4FGL-DR2 catalogue. 
The results for the performance measures are reported for both the DNN and RNN benchmark architectures as described in Sec.~\ref{sec:method}. 
We first of all note that the DNN and RNN perform similar on the task of classifying AGN vs.\ PSR. 
We recall that our benchmark architectures use the energy and time spectra, and the corresponding significances, for each source.
We also test some variations, where one of the two  input spectra is substituted by standard features in the catalogue, see Sec.~\ref{sec:data}.

\begin{table*}
\caption{Performance of AGN vs.\ PSR classification using the 4FGL-DR2 catalogue. For each architecture (DNN or RNN) we report the value of the performance measures defined in Sec.~\ref{subsec:measures}, in particular Eq.~\ref{Eq:crossentropy} for the cross entropy. The  uncertainties of the performance measures combine those from different network initialisations as well as from the selected training data, see Sec.~\ref{subsec:training}.}
\centering
\begin{tabular}{ c @{\hspace{10px}} c @{\hspace{10px}} c @{\hspace{10px}} c @{\hspace{10px}} c @{\hspace{10px}} c @{\hspace{10px}} c} 
\hline\hline

Architecture	& Accuracy [\%]		& F1 score [\%]	& Equal acc	[\%]	& AUC [\%] & Cross entropy  \\ \hline
DNN				& $97.59 \pm 0.35$	& $81.73 \pm 3.43$			& $95.67 \pm 0.83$	& $97.99 \pm 0.75$	& $0.089 \pm 0.019$	\\ \hline
RNN				& $97.31 \pm 0.36$	& $79.27 \pm 2.82$			& $93.86 \pm 1.53$	& $97.40 \pm 0.89$	& $0.089 \pm 0.012$	\\
\hline\hline
\end{tabular}
\label{tab::results4FGL}
\end{table*}

Using the energy spectrum in combination with the Variability Index feature instead of the time series data, we obtain for the DNN an accuracy of $97.90 \pm 0.44$\%, a F1 score of $84.09 \pm 3.27$\% 
and a cross entropy of  $0.079 \pm 0.023$. Using only the energy spectrum data and no additional information about the time dependence, we reach a similar performance (accuracy of $97.70 \pm 0.32$\%, a F1~score of $82.30 \pm 2.48$\% and a cross entropy of  $0.079 \pm 0.015$). When using only the time series, we observe a decrease in performance to an accuracy of $95.46 \pm 0.33$\%, a F1~score of $64.04 \pm 3.92$\% and a cross entropy of  $0.136 \pm 0.01$. 
We conclude that the essential information for the AGN vs.\ PSR classification is contained in the energy spectrum integrated over time. The time series data does not provide significant additional information. The same is observed for the RNN architecture. 

\begin{figure}
\centering
\includegraphics[width=.48\textwidth]{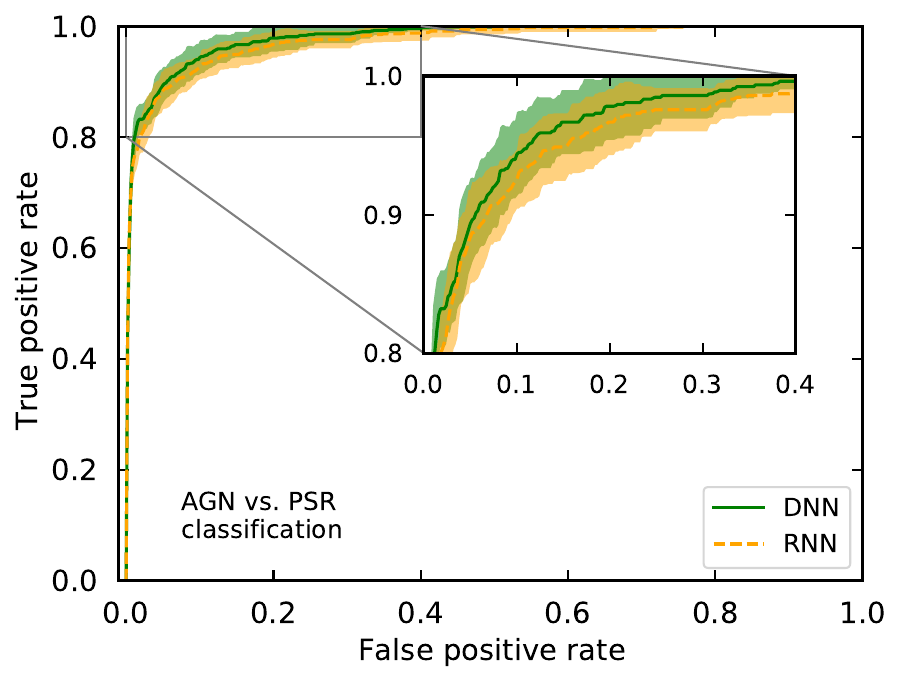}
\includegraphics[width=.48\textwidth]{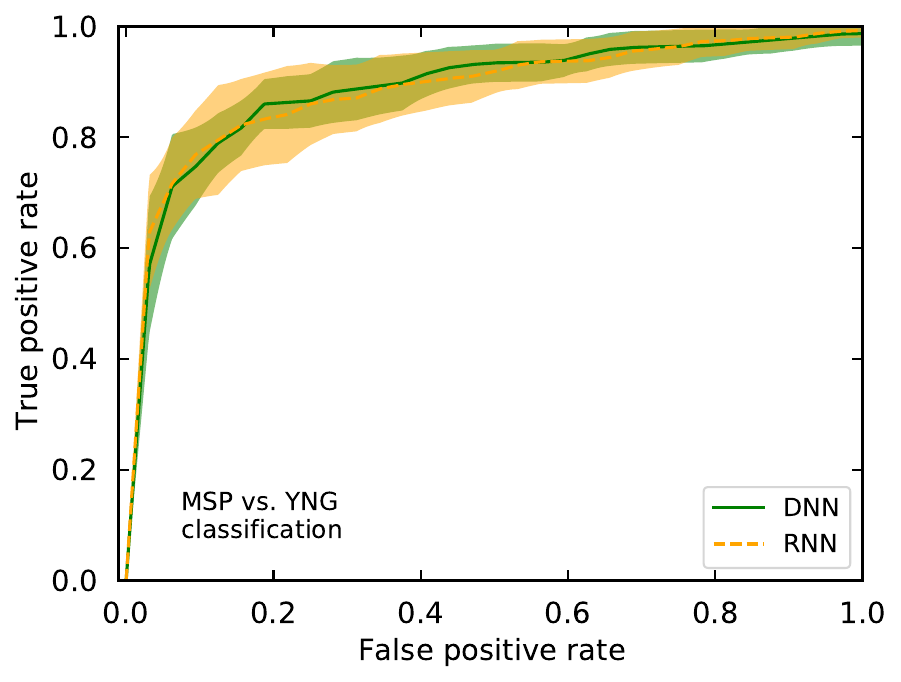}
\caption{ROC curves for AGN vs.\ PSR classification (top) and MSP vs.\ YNG classification (bottom) for the DNN (green solid line) and for the RNN (dashed orange line). The shaded area is given by the standard deviation for the ten runs at each position. 
\label{fig:roc}} 
\end{figure}

We show the ROC curves for the DNN and RNN in the top panel of Fig.~\ref{fig:roc}. The DNN performs slightly better for all PSR efficiencies. Overall, the curves are close to the step function corresponding to an ideal classifier, and the fluctuations, denoted by the coloured bands, are rather small.

In Table~\ref{tab::confmatr} we report the confusion matrix for both RNN and DNN. The numbers presented are the mean and average of ten trainings with new initialisations and different training and testing sets. The very similar numbers reported for the DNN and RNN (in particular for the True AGN) are a consequence of their similar accuracies, but do not necessarily count the same set of sources. 
In fact, while the size of the test set is always the same (1053 AGN and 78 PSR), the individual sources differ when performing the ten training and testing splits before computing the mean. 
One can normalise the entries of the confusion matrix by the sum over the columns to obtain intra-class accuracies. Both DNN and RNN classify $\sim$99\% of AGN correctly, but differ in the numbers of correctly classified PSR. While the RNN has fewer PSR correctly classified ($\sim$74\% compared to $\sim$79\% for the DNN), it is more stable with respect to this class, as can be seen by the smaller fluctuations. Alternatively, one can normalise row-wise and obtain the accuracy given a predicted class. Due to the higher number of falsely classified PSR by the RNN, the DNN's accuracy when having predicted a PSR is slightly higher ($\sim$84.9\% compared to $\sim$84.0\%).

\begin{table}
\caption{Confusion matrix for DNN and RNN classifying AGN vs.~PSR. Values correspond to the mean of the absolute number of sources in the training set. The estimated uncertainty corresponds to the standard deviation of the corresponding entry for ten training runs with new initialisations  and different random splits between training and test data.}
\center
\begin{tabular}{c|c|c|c}
	\hline \hline
	- 								& Network	& True AGN 		& True PSR 		\\ \hline
	\multirow{2}{*}{Predicted AGN}	& DNN		& $1042.1 \pm 3.7$	& $16.4 \pm 5.4$ 	\\ 
									& RNN		& $1042.3 \pm 4.3$	& $19.6 \pm 3.3$	\\ \hline
	\multirow{2}{*}{Predicted PSR}	& DNN		& $10.9 \pm 3.7$ 	& $61.6 \pm 5.4$ 	\\
									& RNN		& $10.7 \pm 4.3$	& $58.4 \pm 3.3$	\\ \hline \hline
\end{tabular}\label{tab::confmatr}
\end{table}

Removing the significances for both series gives the same results for the mean values of our performance metrics, while  their deviations slightly decrease: accuracy $97.60 \pm 0.20$ ($97.67 \pm 0.24$), F1 score $81.68 \pm 1.62$ ($81.82 \pm 2.36$), cross entropy  $0.104 \pm 0.027$ ($0.079 \pm 0.011$) for the DNN (RNN). This originates from the fewer inputs when not using the significances. Since  the significances do not contribute to the overall class separation, they can be seen as additional noise leading to some overfitting and thus a stronger variation on the test sets. We nevertheless keep the significance as input in our standard set-up, as it provides valuable information for the further classification of the PSR class into MSP and YNG, see Sec.~\ref{sec:mspyng}.

 \citet{Luo_2020} state an accuracy of $99.19 \pm 0.16$\% for the AGN vs.\ PSR classification using automated feature selection and a random forest classifier on the 4FGL sources. While \citet{Luo_2020} do not provide any other performance measures, we achieve comparable results on the accuracy, without the need to extract and select features from the catalogue.  

One way to address the imbalance of the data set is by weighting the loss of the AGN and PSR classes such that the corresponding contributions to the total loss are equal. We have also explored techniques for augmenting data of the subdominant PSR class. First, we simply reused the actual PSR data as often as necessary to balance the two classes. Second, we reused the PSR data but added some Gaussian noise to the augmented data points. Third, we adopted the SMOTE algorithm~\citep{chawla2002smote}, which can be formulated as $S_{\rm aug} = S_1 + r (S_2 - S_1)$: we pick two PSR sources randomly and add to the first source $S_1$ the difference to the second source $S_2$, multiplied by a random number $r$, drawn from a uniform distribution in the interval $(0, 1)$. The difference is taken for each input that we use individually. We thus obtain an augmented source $S_{\rm aug}$ with spectra interpolating between those of the two original sources. Overall, we find that these different data augmentation techniques do not improve the classification. 
While the performance at a generic threshold of 0.5 changes, i.e.\ the accuracy in the PSR and AGN classes increases and decreases, respectively, these effects can be compensated for by a change in the decision threshold. Since we adapt this decision threshold according to the performance in the training set, we achieve no benefit from using the above-mentioned data augmentation techniques.

\subsubsection{MSP vs.\ YNG classification}\label{sec:mspyng}
The classification of MSP vs.\ YNG pulsars is significantly more difficult, given the very small data set (only 230 sources, with 127 YNG and 103 MSP, of which we use 161 for training). Keeping the batch size as in Sec.~\ref{sec:agnpsr}, the network parameters are updated only few times during one epoch (one iteration through all training sources). A further decrease of the batch size leads to unstable estimates of the gradients during training and should therefore be avoided. We thus train the networks for more epochs for the task of MSP vs.\ YNG classification (150 instead of 50).
Additionally to the energy spectrum and time series information, the latitude of the pulsar in Galactic 
coordinates (\texttt{GLAT}) is a distinguishing feature that can be directly measured, and it has been used in previous attempts to differentiate between these two source classes \citep{Luo_2020, Saz_Parkinson_2016}. 
Given our flexible neural network architectures, we can easily include such features in the classification. 

In Table~\ref{tab::resultsMSPvsYNG} we report our results for the MSP vs.\ YNG classification task. We find that adding the latitude improves the different performance measures by about 3\% for the DNN and 4\% for the RNN. 
We also observe that removing the time series has no strong effect on the classification power in terms of accuracy and F1 score, but increases the cross entropy and its fluctuations slightly for the RNN and more significantly for the DNN. The overall large fluctuations within the cross entropy highlight once more the difficulty of convergent and stable training with such a small dataset. Here, the RNN shows some advantage over the DNN.
In contrast to the AGN vs.\ PSR classification, we see a decrease in performance when leaving out the significance. Both, DNN and RNN, drop below 80\% accuracy and 0.9 AUC.
To be compared with our results, the feature selection algorithm from \citet{Luo_2020} achieves $89.61 \pm 2.34$\% for the MSP vs.\ YNG classification task using a boosted logistic regression as classifier.


\begin{table*}
\caption{Same as Table \ref{tab::results4FGL} but for the performance of MSP vs.\ YNG classification using the 4FGL-DR2 catalogue.}
\begin{tabular}{ c  c c c c c c} 
\hline\hline

Architecture			& Inputs				& Accuracy [\%]		& F1 score [\%]	& Equal acc	[\%]	& AUC [\%]	& Cross entropy  \\ \hline
\multirow{3}{*}{DNN}	& History, Band 		& $83.91 \pm 3.69$	& $84.47 \pm 3.50$			& $83.91 \pm 3.97$	& $90.25 \pm 2.65$	&  $0.640 \pm 0.278$	\\ 
						& History, Band, GLAT	& $87.25 \pm 3.82$ 	& $88.00 \pm 3.66$			& $86.67 \pm 4.14$	& $93.34 \pm 2.09$	&  $0.611 \pm 0.207$	\\
						& Band, GLAT			& $87.10 \pm 3.86$ 	& $87.82 \pm 3.63$			& $86.67 \pm 4.84$	& $92.84 \pm 3.41$	& $0.689 \pm 0.447$	\\ \hline
\multirow{3}{*}{RNN	}	& History, Band			& $82.03 \pm 5.15$	& $82.69 \pm 4.50$			& $83.33 \pm 5.11$	& $90.03 \pm 4.11$	&  $0.401 \pm 0.086$ 	\\
						& History, Band, GLAT	& $86.52 \pm 2.83$	& $87.46 \pm 2.75$			& $84.78 \pm 2.36$	& $93.04 \pm 2.32$	&  $0.362 \pm 0.057$	\\
						& Band, GLAT			& $85.51 \pm 3.37$	& $85.72 \pm 3.90$			& $85.51 \pm 4.00$	& $92.61 \pm 2.83$	&  $0.375 \pm 0.081$	\\
\hline\hline
\end{tabular}
\label{tab::resultsMSPvsYNG}
\end{table*}

In the bottom panel of Fig.~\ref{fig:roc} we show the ROC curves for the classification of MSP vs.\ YNG using our standard architectures (without Galactic latitude). The performances of the DNN and RNN are very similar over the full range of false positive rates. The comparison to the AGN vs.\ PSR classification (top panel of Fig.~\ref{fig:roc}) shows that the sub-classification task is harder. The corresponding ROC curves are lower and the uncertainties are larger. One reason for this behaviour is the much smaller amount of labelled sources available for training as well as for evaluation.

\subsection{Cross-match between the 3FGL and 4FGL-DR2 catalogues} \label{sec:xval}

Given the increased gamma-ray statistics, and novel associations based on updated catalogues at other wavelengths, some of the sources labelled as UNC in the previous 3FGL catalogue have now a well-defined association in the 4FGL-DR2 catalogue. In total, 294 unclassified sources in the 3FGL catalogue are now labelled as AGN (258) or PSR (36) in the 4FGL-DR2 catalogue.
These sources constitute a cross-match test set to evaluate the capability of our classifiers, by training on the 3FGL catalogue and testing the classification for the previously UNC objects against the classification in the 4FGL-DR2 catalogue. 
In addition, the cross-match test data set allows for an estimate of a potential sample selection bias (also called covariate shift or Malmquist bias in astronomy), see \citet{2019A&A...627A..21P,Luo_2020}.
Data sets in astrophysics are typically based on brightness-limited surveys. Thus the training data sets in supervised classification tasks may be biased towards brighter sources, or sources easier to detect in general, while the classification algorithm is then used to predict the properties of fainter objects observed in more recent surveys. 
Within the context of our work, the sample of bright AGN and PSR labelled in a given catalogue might not necessarily reflect the characteristics of dim sources, which are the dominant part of the UNC sources we want to classify (see e.g.\ 
Fig.~\ref{fig:catstats}). This effect could bias our classifier towards the characteristics of the brighter sources. With this cross-match data set  we can test the performance of our classifier to predict the actual labels of the UNC sources. 

Additionally, the cross-matching between the 3FGL and 4FGL-DR2 catalogues allows us to evaluate the behaviour of our classifier with an increasing amount of training data. We train 
ten networks for each architecture on 60, 70, 80, 90 and 100 percent of the labelled sources in the 3FGL catalogue, and use the cross-matched data set of the 4FGL-DR2 catalogue as the test set. 

Note that the data format of the 3FGL catalogue is somewhat different from the 4FGL-DR2 catalogue as detailed in Sec.~\ref{sec:data}.

In Table~\ref{tab::crossmatch} we report the accuracy, the AUC and the cross entropy of the 3FGL and 4FGL-DR2 catalogue cross-matching (with PSR as positive class). The error corresponds again to the standard deviation of the ten trainings. The accuracies of our neural network classifiers are comparable to the accuracies stated in \citet{Luo_2020} for various machine learning algorithms trained on selected features. By using 100 percent of the sources in the 3FGL catalogue to train our classifiers, we achieve an accuracy of $94.73\pm0.46$  with the RNN, and very similar values for the DNN. This accuracy is lower with respect to the one we find when using the 4FGL-DR2 catalogue for training and testing (accuracies of $97.59 \pm 0.35$\% and $97.31 \pm 0.36$\% for the DNN and RNN, respectively, see Table~\ref{tab::results4FGL}), or if we compute the accuracy within the 3FGL catalogue without using the cross-match sources. In fact, for the training on 70\% of 3FGL sources we get an accuracy of $96.99 \pm 0.50$\% when we test on the remaining 30\% of labelled sources. The same is true for the cross entropy, where we find  $0.123 \pm 0.046$ for the DNN, i.e.\ a factor two better than on the unidentified sources.
This is likely due to the sample selection bias, as the training set is biased towards the brightest sources, while the cross-match set is composed of dim sources, which might not follow exactly the same distribution in feature space. 
A possible way to address this issue using features is presented in Sec.~\ref{sec:autoenc}.

\begin{table*}
\caption{Results for the cross-match classification of the 3FGL and 4FGL-DR2 catalogues. For each architecture, we report the accuracy, the AUC and the cross entropy (see Sec.\ref{subsec:measures}) obtained using 70\% and  100\% of the sources  in the 3FGL catalogue for training.}
\centering
\begin{tabular}{ c @{\hspace{10px}} c @{\hspace{10px}} c  @{\hspace{10px}} c} \hline\hline

\textbf{Architecture}	& Measure	& 70\% Training set	&  100\% Training set	\\ \hline
\multirow{3}{*}{DNN}	& Acc [\%]	&  $94.63 \pm 0.42$	&   $94.42 \pm 0.44$	\\
						& AUC [\%]	& $92.69 \pm 2.31$ 	&  $92.96 \pm 1.48$ 	\\ 
						& Cross entropy 	&  $0.249 \pm 0.046$ &  $0.248 \pm 0.055$	\\ \hline
\multirow{3}{*}{RNN}	& Acc [\%]	&  $94.05 \pm 0.67$	&  $94.73 \pm 0.46$	\\
						& AUC [\%] &  $94.84 \pm 1.76$ &  $95.13 \pm 0.77$ \\
						& Cross entropy 	&  $0.188 \pm 0.026$	&  $0.183 \pm 0.020$ \\
\hline\hline
\end{tabular}
\label{tab::crossmatch}
\end{table*}

We do not observe a significant increase in performance when increasing the training set. 
This is somewhat surprising, as we expect a stronger dependence of the performance of our neural networks on the amount of training data than in the case of the standard machine learning algorithms used in previous analyses. 
It is likely that larger increases of the amount of training data are needed in order to observe a significant trend in the performance within the uncertainties.  

\subsection{Predictions for UNC sources in the 4FGL-DR2 catalogue} \label{sec:pred}
We finally perform a two-step classification of the UNC sources in the 4FGL-DR2 catalogue. We first classify sources as either AGN or PSR, and then classify the corresponding PSR sources  as either MSP or YNG.  We specify that a source is classified as PSR with confidence if the mean prediction of the ten classifiers (obtained from ten training runs, see Sec.~\ref{sec:results}) for the PSR class is above 0.9.  The same holds for the AGN classification. In Table~\ref{tab::UNIDpred} we show the number of classified sources and an estimated accuracy. 
These  are estimated as mean and standard deviation of the accuracies obtained on the test sets for the ten classifiers. 
Results in Table~\ref{tab::UNIDpred} are reported for six different architectures (described in Sec.~\ref{sec:mspyng}): DNN (RNN), DNN (RNN) adding the Galactic latitude for the sub-classification of YNG versus MSP, and DNN (RNN) Band GLAT, where we removed the time series data and use only the energy spectrum data in combination with the Galactic latitude.
As discussed in Sec.~\ref{sec:mspyng}, adding the latitude increases the performance in the MSP vs.\ YNG classification. In Table~\ref{tab::UNIDpred} we illustrate this effect for the classification of UNC sources. 
Since we add the latitude only for the sub-classification of the PSR sources, the classification of the UNC sources in terms of AGN and PSR is not affected.  

\begin{table*}
\caption{Number of classified sources obtained for each  class for different DNN  and RNN  architectures, respectively. Shown are results using both energy and time series (DNN/RNN), adding the Galactic latitude (DNN/RNN GLAT), and leaving out the time series (DNN/RNN Band GLAT). The quantity in brackets corresponds to an accuracy estimate obtained from the performance on the labelled test sets}.
\label{tab::UNIDpred} 
\begin{tabular}{c c c c c}

\hline \hline
	Architecture	& PSR (acc)				& AGN (acc)					& YNG (acc)				& MSP (acc) \\ \hline
	DNN				& 78 ($92.75 \pm 3.25$)	& 1050 ($98.91 \pm 0.32$)	& 22 ($93.70 \pm 5.20$)	& 21 ($86.14 \pm 4.79$) \\
	DNN GLAT		&						&							& 20 ($92.19 \pm 5.26$)	& 22 ($89.49 \pm 4.49$) \\
	DNN Band GLAT	&						&							& 21 ($91.45 \pm 5.66$)	& 24 ($90.17 \pm 4.93$) \\\hline
	RNN 			& 57 ($94.14 \pm 2.64$)	& 1076 ($98.86 \pm 0.27$)	& 5 ($97.94 \pm 2.73$)	& 0 (--) \\
	RNN GLAT		&						&							& 2 ($97.07 \pm 2.79$)	& 12 ($95.80 \pm 5.81$) \\
	RNN Band GLAT	&						&							& 2 ($97.68 \pm 4.61$)	& 8 ($93.55 \pm 6.71$) \\
\hline \hline
\end{tabular}
\end{table*}

\begin{figure*}
\centering
\includegraphics[clip, width=.75\textwidth]{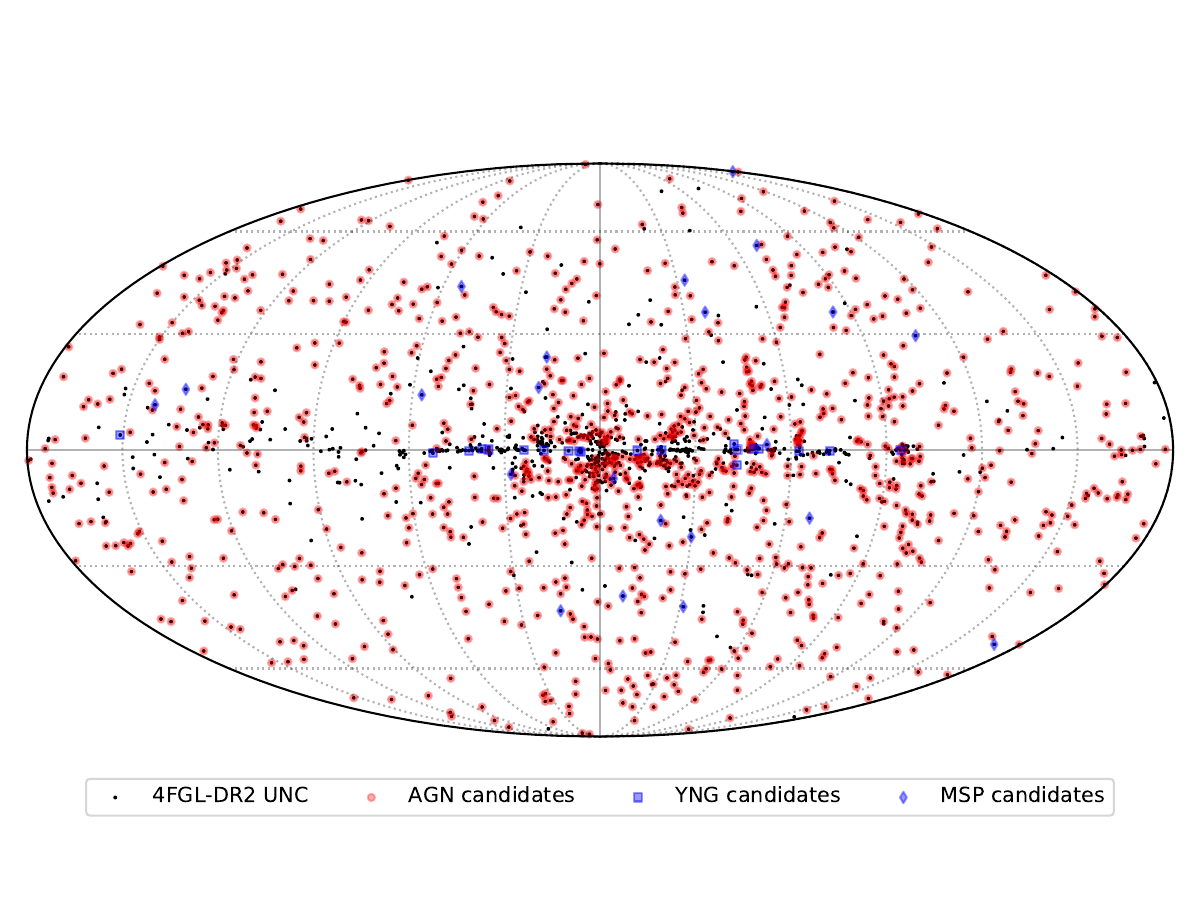}
\caption{Sky distribution in Galactic coordinates of 4FGL-DR2 unclassified sources (black dots). AGN candidates identified by our DNN classifier are additionally marked with red circles, while blue squares and diamonds indicate the YNG pulsar and MSP candidates identified using our DNN together with the Galactic latitude, respectively.  \label{fig:map_classified}}
\end{figure*}

\begin{figure*}
\includegraphics[clip, width=.49\textwidth]{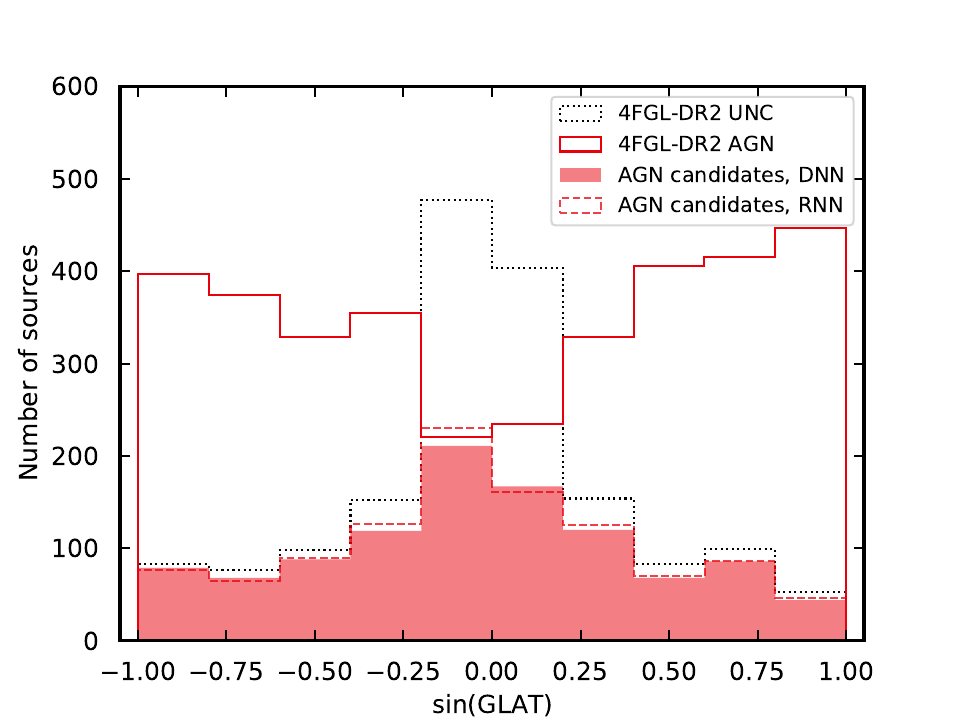}
\includegraphics[clip, width=.49\textwidth]{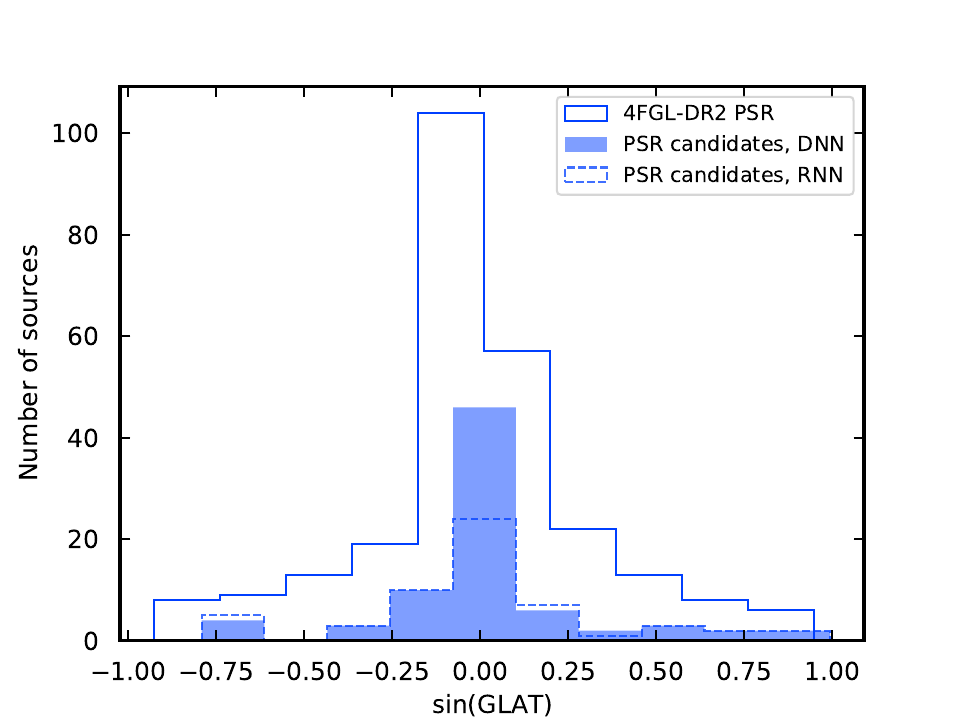}
\caption{Galactic latitude distribution of AGN (left panel) and PSR (right panel) of 4FGL-DR2 sources. 
In each panel, the distribution of AGN/PSR labelled sources in the 4FGL-DR2 is reported together with the distribution of source candidates as derived by our two neural network architectures. The distribution of UNC sources is also shown in the left panel for comparison.
\label{fig:hist_classified}}
\end{figure*}

The sky distribution of AGN, YNG and MSP candidates obtained with the DNN classifier, together with the UNC sources in the 4FGL-DR2 catalogue, is presented in Fig.~\ref{fig:map_classified}.
The YNG and MSP sources in Fig.~\ref{fig:map_classified} are obtained by adding the Galactic latitude as input to the classification.
The map shows the position of the 4FGL-DR2 UNC sources, together with the candidate AGN and PSR sources, in Galactic coordinates in a Mollview projection. 
A complementary view of their latitude distribution is presented in Fig.~\ref{fig:hist_classified}. In the left (right) panel we show the AGN (PSR) Galactic latitude distribution  for labelled (associated and identified) sources in the catalogue together with the distribution of source candidates found by our two neural network architectures. The distribution of UNC sources in the 4FGL-DR2 catalogue is also reported in the left panel for comparison.
As expected from the observed distribution of labelled sources, PSR candidates are clustered along the Galactic plane. AGN candidates are distributed more isotropically, although slightly peaked at low latitudes where the majority of UNC sources is located. 
This is compatible with an isotropic distribution of AGN in the inner Galaxy. In fact, by summing the labelled and the candidates sources, we find a similar value of about 350 to 400 AGN for each bin of sin(GLAT)$\,<|0.25|$.
Moreover, YNG pulsar candidates are concentrated at low latitudes, while MSP candidates can be found also at high latitudes away from the Galactic plane.


The DNN classifier predicts that among the 1679 UNC sources in the 4FGL-DR2 catalogue, 1050 are candidates for AGN ($62$\%), 78 for PSR ($4.6$\%), while 551 remain unclassified. These 551 sources are mostly concentrated along the Galactic plane (see Fig.~\ref{fig:map_classified}), where the bright Galactic emission likely prevents a precise measurement of their energy and time spectra. Indeed, more than half of the 551 sources that remain unclassified have a low significance ($\sigma$=5-10) and have been marked with some \texttt{Flags}\footnote{Sources have non-zero analysis flags when the systematic uncertainties of features depend strongly on the uncertainties of modelling the diffuse gamma-ray background, or in the case of low source-to-background ratio, or in the case of source confusion etc., see \citet{Fermi-LAT:2019yla}.}. Similar results are found by the RNN, with 1076 AGN candidates ($64$\%), 57 PSR candidates ($3.4$\%), and 546 sources remaining unclassified.
We find that the RNN classifies a smaller number of PSR candidate sources compared to the DNN, however with higher estimated accuracy. Without adding the latitude information, the RNN is not able to find any MSP, because the separation of the two classes by the classifier is not strong enough to pass the threshold.
The percentage of AGN and PSR candidates is in line with similar estimates in previous publications  \citep{Doert_2014,Saz_Parkinson_2016}, and in fair agreement with the distribution among labelled sources in the 4FGL-DR2 catalogue ($61$\% AGN and $4.5$\% PSR). The predicted labels for UNC sources obtained with the two architectures are overall consistent: for 91.4\% of the sources classified as AGN or PSR by the DNN, the RNN obtains the same classification. Note that the RNN finds only 57 PSR, and the score is thus limited to 98.1\%. While the two classifiers differ in the certainty of their predictions, they still give consistent results. If we modify the threshold of the RNN to a lower value of 0.8 (0.7) we obtain an agreement of 97.7\% (99.2\%). In fact, we verified that all sources identified as PSR or AGN by the DNN are also classified to belong to the same class by the RNN when reducing the threshold to 0.5. 

\begin{figure*}
\centering
\includegraphics[clip, width=.6\textwidth]{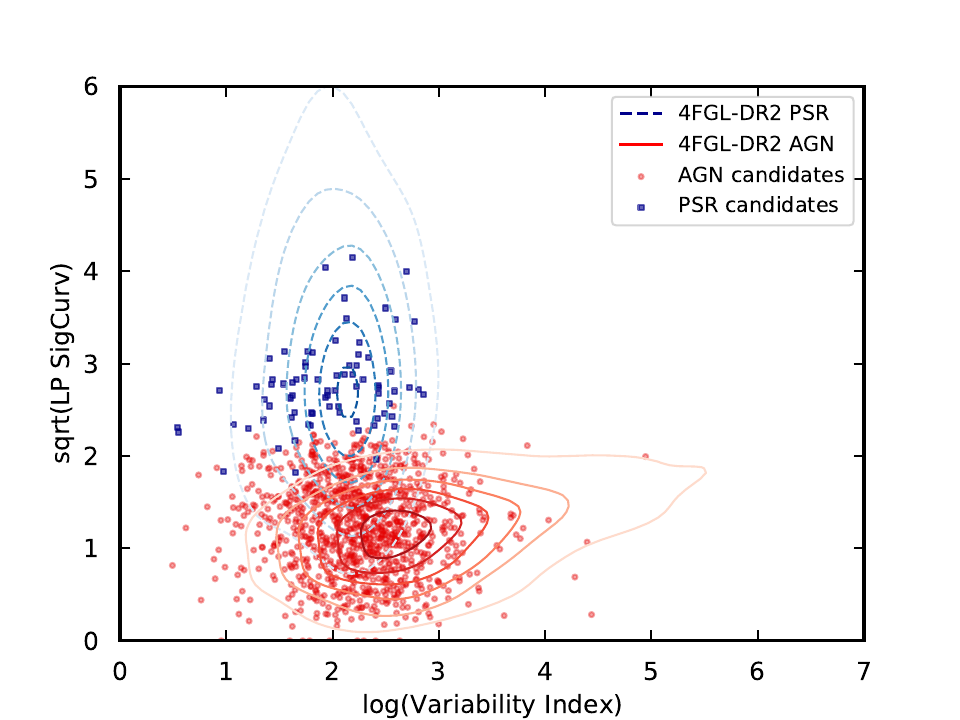}
\caption{Curvature-Variability plot (Sqrt(LP SignCurv) vs.\ Log(Variability Index)) for labelled  AGN (red contours) and PSR (blue dashed contours) sources in the 4FGL-DR2 catalogue. The distribution of candidate AGN (red circles) and PSR (blue squared) obtained from the DNN is superimposed.  \label{fig:cont_classified}}
\end{figure*}

Although our DNN and RNN architectures use only the energy and time spectra for training, testing, and in order to provide a prediction for the UNC sources, the distribution of relevant features can be analysed afterwards. 
In Fig.~\ref{fig:cont_classified} we show a variability-curvature plot for the labelled sources in the 4FGL-DR2 catalogue, obtained by extracting the variability index (\texttt{Variability\_index}) and  the curvature significance for the log parabola spectral fit (\texttt{LP\_SignCurv}). The red contours refer to AGN and the blue dashed contours to PSR. In the variability-curvature plane AGN and PSR are known to cluster in different regions, as  PSR tend to be non-variable on long time scales and have spectra exhibiting significant curvature (see e.g.\ \citet{2010ApJS..188..405A}). 
In Fig.~\ref{fig:cont_classified} we also show the distribution of candidate AGN (red circles) and PSR (blue squares) sources identified by the DNN. Even though the DNN is not trained to separate sources based on these features (as done instead when using feature selection, see e.g.\ \citet{Saz_Parkinson_2016,Luo_2020}), the distribution of candidate sources follows the expected clustering in the variability-curvature plane, confirming that the neural networks extract the relevant features from the energy and time spectra. Similar results are obtained for the RNN. 

The separation of the two classes from these features looks very straightforward. However, the sources included in Fig.~\ref{fig:cont_classified} are classified with high certainty,  and a clear separation is thus expected. Moreover, we chose the features specifically to illustrate such a separation. For other features the separations are not as obvious.

\begin{table*}
\caption{List of MSP candidates selected from the UNC sources in the 4FGL-DR2 catalogue by our DNN and RNN GLAT architectures. The last two columns contain a checkmark if the source is found also in the MSP candidate list of \citet{Saz_Parkinson_2016} (SP16) or \citet{Luo_2020} (L20). \label{tab::msp}}
\begin{tabular}{c c c c c c }
\hline \hline
Source name	& GLON [deg]				& GLAT [deg]				& Architecture		&  SP16 & L20  \\ \hline
	4FGL J0312.1-0921  & 191.496 & -53.37 & DNN/RNN &  $\checkmark$   &           - \\
4FGL J0953.6-1509 & 251.903 & 29.6028  & DNN/RNN &   $\checkmark$  & -  \\
4FGL J1120.0-2204 & 276.492 & 36.0559  & DNN/RNN &   $\checkmark$  & - \\
4FGL J1221.4-0634 & 289.711 & 55.5178  & DNN/RNN &   - & - \\
4FGL J1225.9+2951 & 185.427 & 83.7729   & DNN/RNN &   - &  - \\
4FGL J1400.0-2415 & 322.353 & 36.0036  & DNN/RNN &   - &  -\\
4FGL J1400.6-1432 &  326.982& 45.0692 &  DNN/RNN &  -  &  $\checkmark$ \\
4FGL J1627.7+3219 & 52.9838 & 43.2368 &  DNN/RNN &  $\checkmark$   & -\\
4FGL J2043.9-4802 & 351.629 &-38.2837  &  DNN/RNN &  -  & - \\
4FGL J2112.5-3043 &  14.9039& -42.4435 &  DNN/RNN &  $\checkmark$     &          $\checkmark$ \\
4FGL J2133.1-6432 & 328.748 & -41.2888 &  DNN/RNN &  $\checkmark$&- \\
4FGL J2212.4+0708 & 68.7857 & -38.4799 &  RNN     &  $\checkmark$ &              $\checkmark$  
\\ \hline \hline
\end{tabular}
\end{table*}

Let us now discuss the classification of MSP vs.\ YNG pulsars in more detail. Among the  21  sources classified as MSP by our standard DNN, 16 are contained within the 22 and 24 MSP sources found when adding the latitude information and only using the energy spectrum data, respectively.
In total, these three different classifications give 29  MSP candidates. 
As for the RNN, the eight sources classified as MSP using the energy spectrum data in combination with the Galactic latitude GLAT are part of the 12 sources classified using RNN with energy, time spectra  and GLAT; 11 of these are included in those found by the DNN.   
Considering the YNG class, the five candidate sources found by the RNN are all contained within the YNG candidate sources obtained by the DNN. 

In Table~\ref{tab::msp} we provide a list of 11 promising  MSP candidates identified by both our DNN GLAT and RNN GLAT architectures, plus an additional source found with high certainty by the RNN only. The last two columns illustrate the cross-match of our list with previous analyses searching for MSP candidates \citep{Saz_Parkinson_2016,Luo_2020}.

An overlap of seven out of 13 MSP candidates is found with the list provided in \citet{Saz_Parkinson_2016}. 
The sources 4FGL~J1221.4-0634 (3FGL~J1221.5-0632), 4FGL J1225.9+2951 (3FGL J1225.9+2953), 4FGL J1400.0-2415 (3FGL J1400.2-2413), 4FGL J2043.9-4802 (3FGL J2043.8-4801) and 4FGL~1400.6-1432  (3FGL 1400.5-1437)
are not present in the list of high-confidence candidates in \citet{Saz_Parkinson_2016}, most probably since these sources had a low significance in the 3FGL; because of the low significance, they have not been classified as probable AGN. 
Our overlap with the list provided in \citet{Luo_2020} is smaller, and only three sources are found in both lists. 
Since the two lists are drawn according to different classification thresholds to define high-confidence candidates, it is likely that a larger overlap could be found taking more similar cuts. 
The first two and the last three sources in Table~\ref{tab::msp} are also found in the list of Galactic candidates provided by \citet{Mirabal_2016}. 
Future multiwavelength observation campaigns targeted towards these candidate sources are required to finally clarify their association with MSP, YNG or other objects. 
The full list of the DNN and RNN predictions for the classes of UNC sources in the 4FGL-DR2 catalogue is available at \url{https://github.com/manconi/agn-psr-nn-classification}. 

In order to check the robustness of our MSP vs.\ YNG classifiers we performed a further test. Instead of using the sources classified as PSR as a starting point for our YNG vs.\ MSP classification, we start with the UNC sources classified as AGN. 
The DNN tends to label these sources as mostly MSP (16 YNG, 414 MSP and 621 below threshold). 
This underlines the need for a proper definition of an uncertainty for each classified source. This could for example be achieved by transforming our classifiers into Bayesian neural networks, which  can provide an uncertainty for each source and might result in high uncertainties for classes that are not used for training, see e.g.\ \citet{2020MNRAS.491.4277M}.  We leave this extension to future work.

\section{Feature selection based on autoencoders} \label{sec:autoenc}

As  argued in Sec.~\ref{sec:xval}, when the distributions of the training and the test data sets do not match,  a sample selection bias might reduce the classifier performance.  We demonstrate in this section that  feature selection with an autoencoder does not rely on the source labels, and thus may help to reduce the problem of a sample selection bias.

One of the main problems when dealing with classification using machine learning on small datasets 
is the tendency for overfitting of the training data, resulting in bad generalisation to unseen data.\footnote{We indeed observe overfitting also for our standard architectures, especially when looking at the accuracy in the PSR class or the F1 score. Our standard DNN gives an F1 score of 81.73\% on the test data and 92.28\% on the training data. Similar results are obtained for the RNN.} Feature selection reduces the number of inputs and can prevent the classification algorithm to learn training-set specific information. In~\citet{Luo_2020} the authors apply a recursive feature elimination procedure using a Random Forest (RF) algorithm as a classifier. In this setup a RF classifier is trained on labelled data, and each feature is assigned an importance score. The feature with the lowest score is eliminated and another RF classifier is trained on the new subset. This procedure is repeated until a stopping condition is fulfilled, which can either be a number of remaining features or, as in~\citet{Luo_2020}, a performance ratio between two consecutive classifiers.

While this algorithm is selecting the best features for classification, it might be affected by the sample selection bias (covariate shift). In fact, the algorithm relies on labelled data, and the selected sample thus corresponds to those sources that are observed with sufficient significance to be classified. 
An alternative way to reduce the number of features used for classification is feature reduction using an autoencoder (AE) \citep{2019A&A...627A..21P}, as we outline below. 

\subsection{Architecture and data prepocessing}
An AE is a neural network architecture that first encodes an input to a reduced, latent dimension and then extracts the original input from this encoding. The latent representation should then contain the most important information about the input, and can thus be used for classification. Since no labels are necessary to train the AE, this is an example of unsupervised learning, and the AE can be trained on either the training set, the testing set or both. By training the AE on the test set one selects features that are most important to describe the test sources. Using these features for training can reduce the sample selection bias. Note that the features are not selected according to the best classification performance, but to encode the most relevant information about the sources.

Our autoencoder consists of three fully connected hidden layers, the second of which provides the encoding. The first and third layer consist of eight nodes and apply PReLU activation. We find that training on the encoding allows a simple fully connected classifier with one hidden layer of 40 nodes to train longer without overfitting. 

Using the source features of the catalogue as input to the AE requires some additional preprocessing. First of all, there are some features that are not measured for all sources. 
We adopt the approach used in \citet{Luo_2020}, where features with missing values in more than five percent of the sources are removed. We further remove those sources where more than five of the remaining features are empty. The remaining missing values are replaced by the mean value of all classified sources. This procedure leaves us with 40 features.

\subsection{Training and testing}
We use 70\% of the remaining sources for training and 30\% for testing. Each feature is normalised such that the mean across the training set is zero and the standard deviation is one. After training the autoencoder on the test set, a reduced set of features is extracted from the latent dimension and used in the final classifier.

To estimate the performance and the uncertainty we train both networks ten times, resampling the training and testing set before each training.

\subsection{Results}
Using the 40 features directly in a shallow classifier with one hidden layer of 40 nodes, we find an accuracy of $97.91 \pm 0.48$\% with an F1-score of $82.52 \pm 3.92$\% (for PSR as positive class). Using the same classifier on a five-dimensional encoding of the features from the AE, we find $97.96 \pm 0.26$\% accuracy and a F1-score of $84.42 \pm 2.04$\%. 

While the overall accuracy is similar, we see that using the AE for feature selection increases the performance stability. This is what we expect from the reduction  of features.
Even if the performance remains  below the performance obtained with a recursive feature elimination procedure based on a RF classifier~\citep{Luo_2020}, we emphasise again that our AE is not trained to select features separating the two classes and might thus be less affected by the sample selection bias. Exploring the potential of autoencoders for automated feature selection in more detail is left for future work. 

\section{Conclusions} \label{sec:theend}
In this paper we have presented a new approach to classify unassociated/unidentified gamma-ray sources with deep neural networks. 
We have focused on the separation between Active Galactic Nuclei (AGN) and Galactic pulsars (PSR), and the sub-classification of pulsars into young (YNG) and millisecond pulsars (MSP) in the last release of the \fermi catalogue of gamma-ray sources, the 4FGL-DR2. Our methods, results and their main novelties are summarized as follows.

The neural networks have been trained directly on the energy and time-dependent photon fluxes as provided in the 4FGL-DR2 \fermi catalogue, without relying on specific, human-crafted features as done in previous work. We have explored two different deep neural networks, a fully connected dense neural network (DNN) and, for the first time in the context of gamma-ray source classification, a recurrent neural network (RNN).
These neural network architectures have been shown to provide  powerful classifiers. They give compatible results, with a performance that is comparable with previous analyses based on human-crafted features.
For our benchmark DNN we reach an accuracy of $97.59 \pm 0.35$\% with a cross-entropy of $0.089\pm 0.019$ for the classification of AGN versus PSR. 
For the sub-classification of pulsars into YNG and MSP,  we obtain an accuracy of $87.25 \pm 3.82$\% with a cross-entropy of $0.611\pm 0.207$ when adding the Galactic latitude.

The accuracy of our classifiers has also been tested with a cross-match procedure, where we have compared our predictions for the unlabelled sources from the previous \fermi 3FGL source catalogue with the classification provided by the current 4FGL-DR2 catalogue. We find an accuracy of $94.42 \pm0.44$\% with a cross-entropy of $0.248\pm 0.055$ when testing against the cross-match data set. 
The lower accuracy with respect to the accuracy found using the 4FGL-DR2 sources for training and testing may point to a sample selection bias, as the training set of the 4FGL-DR2 catalogue may be biased towards the brightest sources.

We have proposed a new method to address the sample selection bias when performing source classification based on features. Using an autoencoder one can select those features for classification that encode the most relevant information about the sources, irrespective of the source labels and of the feature's classification performance. With such a technique we find an accuracy of $97.96\pm 0.26$\% for the classification of AGN versus PSR, comparable to what we found using the energy and time dependent fluxes.

Finally, we have presented our predictions for the classification of unlabelled sources in the 4FGL-DR2 catalogue as either AGN or PSR, and of the PSR sources as candidate YNG or MSP pulsars. 
Considering our benchmark DNN, we identify 1050 candidate AGN and 78 candidate PSR sources. 
The distribution of candidate sources follows the expected Galactic latitude distribution, and most importantly, the clustering in the variability-curvature plane, confirming that neural networks extract the relevant features from the energy and time spectra. We have also provided a list of 12 MSP candidates selected from unlabelled sources and discussed the overlap with results from previous classifications based on other machine learning techniques. 

The sources have been labelled by our deep networks as candidate AGN, PSR, YNG or MSP based on their gamma-ray properties only. They  should thus not be considered identified sources, but rather as sources likely to be associated/identified in the future. We encourage multiwavelength follow-up observations targeting the candidate AGN and PSR, searching for example for radio and gamma-ray pulsations, or spectral variability on larger time scales. 

We emphasise that our method is very flexible and can easily be generalised to include multiwavelength data on the energy and time spectra coming from different observatories, as well as additional selected features. 
For example, spectra can be complemented with data at higher energies coming from forthcoming observations of \fermi sources with CTA \citep{CTAConsortium:2018tzg}. 

Up to now gamma-ray source classification techniques do not provide reliable error estimates for the assignment of sources. Bayesian inference methods, for example, may provide more information about the accuracy and the robustness of the neural networks. We leave the exploration of this aspect to future work.

\section*{Acknowledgements}

 We would like to thank Anja Butter, Mattia Di Mauro, Fiorenza Donato, Alexander M\"uck and Tilman Plehn for insightful discussions. The work of TF is supported by the Research Training Group GRK 2497 of the German Research Foundation DFG.
Computations were performed with computing resources granted by RWTH Aachen University.

\section*{Data Availability}\label{sec:avdata}

The networks used in this paper are implemented using \textsc{Tensorflow 2.1.0} \citep{tensorflow2015-whitepaper} and the built in version of \textsc{Keras} \citep{chollet2015keras}.
The reference dataset is the 4FGL-DR2 \fermi catalogue released on 22 May 2020 (saved in FITS table form in the file named \texttt{gll\_psc\_v23.fit})
available at: \url{https://fermi.gsfc.nasa.gov/ssc/data/access/lat/10yr_catalog/}. 
The public list of LAT-detected gamma-ray pulsars used to build a cross-match between this list and the 4FGL-DR2 catalogue is available at \url{https://confluence.slac.stanford.edu/display/GLAMCOG/Public+List+of+LAT-Detected+Gamma-Ray+Pulsars}. 

The complete list of the neural network predictions for the classes of UNC sources in the 4FGL-DR2 catalogue obtained in this paper is available at \url{https://github.com/manconi/agn-psr-nn-classification}.


\bibliographystyle{mnras}
\bibliography{sample63}

\begin{thebibliography}{}
\makeatletter
\relax
\def\mn@urlcharsother{\let\do\@makeother \do\$\do\&\do\#\do\^\do\_\do\%\do\~}
\def\mn@doi{\begingroup\mn@urlcharsother \@ifnextchar [ {\mn@doi@}
  {\mn@doi@[]}}
\def\mn@doi@[#1]#2{\def\@tempa{#1}\ifx\@tempa\@empty \href
  {http://dx.doi.org/#2} {doi:#2}\else \href {http://dx.doi.org/#2} {#1}\fi
  \endgroup}
\def\mn@eprint#1#2{\mn@eprint@#1:#2::\@nil}
\def\mn@eprint@arXiv#1{\href {http://arxiv.org/abs/#1} {{\tt arXiv:#1}}}
\def\mn@eprint@dblp#1{\href {http://dblp.uni-trier.de/rec/bibtex/#1.xml}
  {dblp:#1}}
\def\mn@eprint@#1:#2:#3:#4\@nil{\def\@tempa {#1}\def\@tempb {#2}\def\@tempc
  {#3}\ifx \@tempc \@empty \let \@tempc \@tempb \let \@tempb \@tempa \fi \ifx
  \@tempb \@empty \def\@tempb {arXiv}\fi \@ifundefined
  {mn@eprint@\@tempb}{\@tempb:\@tempc}{\expandafter \expandafter \csname
  mn@eprint@\@tempb\endcsname \expandafter{\@tempc}}}

\bibitem[\protect\citeauthoryear{Abadi et~al.,}{Abadi
  et~al.}{2015}]{tensorflow2015-whitepaper}
Abadi M.,  et~al., 2015, {TensorFlow}: Large-Scale Machine Learning on
  Heterogeneous Systems, \url {https://www.tensorflow.org/}

\bibitem[\protect\citeauthoryear{{Abdo} et~al.,}{{Abdo}
  et~al.}{2009}]{2009Sci...325..848A}
{Abdo} A.~A.,  et~al., 2009, \mn@doi [Science] {10.1126/science.1176113}, \href
  {https://ui.adsabs.harvard.edu/abs/2009Sci...325..848A} {325, 848}

\bibitem[\protect\citeauthoryear{Abdo et~al.,}{Abdo et~al.}{2010a}]{Abdo_2010}
Abdo A.~A.,  et~al., 2010a, \mn@doi [\apjs] {10.1088/0067-0049/188/2/405}, 188,
  405

\bibitem[\protect\citeauthoryear{{Abdo} et~al.,}{{Abdo}
  et~al.}{2010b}]{2010ApJS..188..405A}
{Abdo} A.~A.,  et~al., 2010b, \mn@doi [\apjs] {10.1088/0067-0049/188/2/405},
  \href {https://ui.adsabs.harvard.edu/abs/2010ApJS..188..405A} {188, 405}

\bibitem[\protect\citeauthoryear{Abdo et~al.}{Abdo
  et~al.}{2013}]{TheFermi-LAT:2013ssa}
Abdo A.,  et~al., 2013, \mn@doi [\apjs] {10.1088/0067-0049/208/2/17}, 208, 17

\bibitem[\protect\citeauthoryear{Abdollahi et~al.}{Abdollahi
  et~al.}{2020}]{Fermi-LAT:2019yla}
Abdollahi S.,  et~al., 2020, \mn@doi [\apjs] {10.3847/1538-4365/ab6bcb}, 247,
  33

\bibitem[\protect\citeauthoryear{Acero et~al.}{Acero
  et~al.}{2015}]{Acero:2015gva}
Acero F.,  et~al., 2015, \mn@doi [\apjs] {10.1088/0067-0049/218/2/23}, 218, 23

\bibitem[\protect\citeauthoryear{{Acero}, {Ackermann}, {Ajello}, {Albert},
  {Baldini}  et~al.}{{Acero} et~al.}{2016}]{2016ApJS..223...26A}
{Acero} F.,  {Ackermann} M.,  {Ajello} M.,  {Albert} A.,  {Baldini} L.,
  et~al., 2016, \mn@doi [\apjs] {10.3847/0067-0049/223/2/26}, \href
  {http://adsabs.harvard.edu/abs/2016ApJS..223...26A} {223, 26}

\bibitem[\protect\citeauthoryear{Acharya et~al.}{Acharya
  et~al.}{2018}]{CTAConsortium:2018tzg}
Acharya B.,  et~al., 2018, {Science with the Cherenkov Telescope Array}.
WSP (\mn@eprint {arXiv} {1709.07997}), \mn@doi{10.1142/10986}

\bibitem[\protect\citeauthoryear{{Ackermann}, {Ajello}, {Albert}, {Allafort},
  {Atwood}  et~al.}{{Ackermann} et~al.}{2012a}]{2012ApJS..203....4A}
{Ackermann} M.,  {Ajello} M.,  {Albert} A.,  {Allafort} A.,  {Atwood} W.~B.,
  et~al., 2012a, \mn@doi [\apjs] {10.1088/0067-0049/203/1/4}, \href
  {http://adsabs.harvard.edu/abs/2012ApJS..203....4A} {203, 4}

\bibitem[\protect\citeauthoryear{{Ackermann}, {Ajello}, {Atwood}, {Baldini},
  {Ballet}  et~al.}{{Ackermann} et~al.}{2012b}]{2012ApJ...750....3A}
{Ackermann} M.,  {Ajello} M.,  {Atwood} W.~B.,  {Baldini} L.,  {Ballet} J.,
  et~al., 2012b, \mn@doi [\apj] {10.1088/0004-637X/750/1/3}, \href
  {http://adsabs.harvard.edu/abs/2012ApJ...750....3A} {750, 3}

\bibitem[\protect\citeauthoryear{{Ackermann}, {Ajello}, {Albert}, {Atwood},
  {Baldini}  et~al.}{{Ackermann} et~al.}{2015}]{2015ApJ...799...86A}
{Ackermann} M.,  {Ajello} M.,  {Albert} A.,  {Atwood} W.~B.,  {Baldini} L.,
  et~al., 2015, \mn@doi [\apj] {10.1088/0004-637X/799/1/86}, \href
  {http://adsabs.harvard.edu/abs/2015ApJ...799...86A} {799, 86}

\bibitem[\protect\citeauthoryear{{Atwood}, {Abdo}, {Ackermann}, {Althouse},
  {Anderson}  et~al.}{{Atwood} et~al.}{2009}]{2009ApJ...697.1071A}
{Atwood} W.~B.,  {Abdo} A.~A.,  {Ackermann} M.,  {Althouse} W.,  {Anderson} B.,
    et~al., 2009, \mn@doi [\apj] {10.1088/0004-637X/697/2/1071}, \href
  {http://adsabs.harvard.edu/abs/2009ApJ...697.1071A} {697, 1071}

\bibitem[\protect\citeauthoryear{Ball \& Brunner}{Ball \&
  Brunner}{2010}]{Ball:2009wd}
Ball N.~M.,  Brunner R.~J.,  2010, \mn@doi [Int. J. Mod. Phys. D]
  {10.1142/S0218271810017160}, 19, 1049

\bibitem[\protect\citeauthoryear{Ballet, Burnett, Digel  \& Lott}{Ballet
  et~al.}{2020}]{Ballet:2020hze}
Ballet J.,  Burnett T.,  Digel S.,   Lott B.,  2020, {Fermi Large Area
  Telescope Fourth Source Catalog Data Release 2} (\mn@eprint {arXiv}
  {2005.11208})

\bibitem[\protect\citeauthoryear{Baron}{Baron}{2019}]{baron2019machine}
Baron D.,  2019, Machine Learning in Astronomy: a practical overview
  (\mn@eprint {arXiv} {1904.07248})

\bibitem[\protect\citeauthoryear{{Becker}, {Pichara}, {Catelan}, {Protopapas},
  {Aguirre}  \& {Nikzat}}{{Becker} et~al.}{2020}]{2020MNRAS.493.2981B}
{Becker} I.,  {Pichara} K.,  {Catelan} M.,  {Protopapas} P.,  {Aguirre} C.,
  {Nikzat} F.,  2020, \mn@doi [\mnras] {10.1093/mnras/staa350}, \href
  {https://ui.adsabs.harvard.edu/abs/2020MNRAS.493.2981B} {493, 2981}

\bibitem[\protect\citeauthoryear{Carleo, Cirac, Cranmer, Daudet, Schuld,
  Tishby, Vogt-Maranto  \& Zdeborov\'{a}}{Carleo et~al.}{2019}]{Carleo_2019}
Carleo G.,  Cirac I.,  Cranmer K.,  Daudet L.,  Schuld M.,  Tishby N.,
  Vogt-Maranto L.,   Zdeborov\'{a} L.,  2019, \mn@doi [Reviews of Modern
  Physics] {10.1103/revmodphys.91.045002}, 91

\bibitem[\protect\citeauthoryear{Chawla, Bowyer, Hall  \& Kegelmeyer}{Chawla
  et~al.}{2002}]{chawla2002smote}
Chawla N.~V.,  Bowyer K.~W.,  Hall L.~O.,   Kegelmeyer W.~P.,  2002, Journal of
  artificial intelligence research, 16, 321

\bibitem[\protect\citeauthoryear{Cheng}{Cheng}{2018}]{Cheng_2018}
Cheng T.,  2018, \mn@doi [Computing and Software for Big Science]
  {10.1007/s41781-018-0007-y}, 2

\bibitem[\protect\citeauthoryear{Chiaro, Salvetti, La~Mura, Giroletti, Thompson
   \& Bastieri}{Chiaro et~al.}{2016}]{Chiaro:2016noj}
Chiaro G.,  Salvetti D.,  La~Mura G.,  Giroletti M.,  Thompson D.,   Bastieri
  D.,  2016, \mn@doi [\mnras] {10.1093/mnras/stw/1830}, 462, 3180

\bibitem[\protect\citeauthoryear{Chollet et~al.}{Chollet
  et~al.}{2015}]{chollet2015keras}
Chollet F.,  et~al., 2015, Keras, \url{https://keras.io}

\bibitem[\protect\citeauthoryear{Doert \& Errando}{Doert \&
  Errando}{2014}]{Doert_2014}
Doert M.,  Errando M.,  2014, \mn@doi [\apj] {10.1088/0004-637x/782/1/41}, 782,
  41

\bibitem[\protect\citeauthoryear{Egan, Fedorko, Lister, Pearkes  \& Gay}{Egan
  et~al.}{2017}]{egan2017long}
Egan S.,  Fedorko W.,  Lister A.,  Pearkes J.,   Gay C.,  2017, Long Short-Term
  Memory (LSTM) networks with jet constituents for boosted top tagging at the
  LHC (\mn@eprint {arXiv} {1711.09059})

\bibitem[\protect\citeauthoryear{Englert, Fairbairn, Spannowsky, Stylianou  \&
  Varma}{Englert et~al.}{2020}]{englert2020sensing}
Englert C.,  Fairbairn M.,  Spannowsky M.,  Stylianou P.,   Varma S.,  2020,
  Sensing Higgs cascade decays through memory (\mn@eprint {arXiv} {2008.08611})

\bibitem[\protect\citeauthoryear{{Fornasa} \& {S{\'a}nchez-Conde}}{{Fornasa} \&
  {S{\'a}nchez-Conde}}{2015}]{2015PhR...598....1F}
{Fornasa} M.,  {S{\'a}nchez-Conde} M.~A.,  2015, \mn@doi [\physrep]
  {10.1016/j.physrep.2015.09.002}, \href
  {http://adsabs.harvard.edu/abs/2015PhR...598....1F} {598, 1}

\bibitem[\protect\citeauthoryear{Fraser \& Schwartz}{Fraser \&
  Schwartz}{2018}]{Fraser_2018}
Fraser K.,  Schwartz M.~D.,  2018, \mn@doi [Journal of High Energy Physics]
  {10.1007/jhep10(2018)093}, 2018

\bibitem[\protect\citeauthoryear{Guest, Collado, Baldi, Hsu, Urban  \&
  Whiteson}{Guest et~al.}{2016}]{Guest_2016}
Guest D.,  Collado J.,  Baldi P.,  Hsu S.-C.,  Urban G.,   Whiteson D.,  2016,
  \mn@doi [Physical Review D] {10.1103/physrevd.94.112002}, 94

\bibitem[\protect\citeauthoryear{{Harding} \& {Muslimov}}{{Harding} \&
  {Muslimov}}{1998}]{1998ApJ...508..328H}
{Harding} A.~K.,  {Muslimov} A.~G.,  1998, \mn@doi [\apj] {10.1086/306394},
  \href {https://ui.adsabs.harvard.edu/abs/1998ApJ...508..328H} {508, 328}

\bibitem[\protect\citeauthoryear{Hinners, Tat  \& Thorp}{Hinners
  et~al.}{2018}]{Hinners_2018}
Hinners T.~A.,  Tat K.,   Thorp R.,  2018, \mn@doi [The Astronomical Journal]
  {10.3847/1538-3881/aac16d}, 156, 7

\bibitem[\protect\citeauthoryear{Hui et~al.,}{Hui et~al.}{2020}]{Hui:2020cmv}
Hui C.,  et~al., 2020, \mn@doi [\mnras] {10.1093/mnras/staa1113}, 495, 1093

\bibitem[\protect\citeauthoryear{{Ishida}}{{Ishida}}{2019}]{2019NatAs...3..680I}
{Ishida} E. E.~O.,  2019, \mn@doi [Nature Astronomy]
  {10.1038/s41550-019-0860-6}, \href
  {https://ui.adsabs.harvard.edu/abs/2019NatAs...3..680I} {3, 680}

\bibitem[\protect\citeauthoryear{Kingma \& Ba}{Kingma \& Ba}{2017}]{adam}
Kingma D.~P.,  Ba J.,  2017, Adam: A Method for Stochastic Optimization,
  Conference paper at the 3rd International Conference for Learning
  Representations, San Diego, 2015 (\mn@eprint {arXiv} {1412.6980})

\bibitem[\protect\citeauthoryear{Kova\v{c}evi\'c, Chiaro, Cutini  \&
  Tosti}{Kova\v{c}evi\'c et~al.}{2020}]{Kovacevic:2020sly}
Kova\v{c}evi\'c M.,  Chiaro G.,  Cutini S.,   Tosti G.,  2020, \mn@doi [\mnras]
  {10.1093/mnras/staa394}, 493, 1926

\bibitem[\protect\citeauthoryear{Louppe, Cho, Becot  \& Cranmer}{Louppe
  et~al.}{2019}]{Louppe_2019}
Louppe G.,  Cho K.,  Becot C.,   Cranmer K.,  2019, \mn@doi [Journal of High
  Energy Physics] {10.1007/jhep01(2019)057}, 2019

\bibitem[\protect\citeauthoryear{Luo, Leung, Hui  \& Li}{Luo
  et~al.}{2020}]{Luo_2020}
Luo S.,  Leung A.~P.,  Hui C.~Y.,   Li K.~L.,  2020, \mn@doi [\mnras]
  {10.1093/mnras/staa166}, 492, 5377

\bibitem[\protect\citeauthoryear{Mirabal, Frias-Martinez, Hassan  \&
  Frias-Martinez}{Mirabal et~al.}{2012}]{Mirabal:2012em}
Mirabal N.,  Frias-Martinez V.,  Hassan T.,   Frias-Martinez E.,  2012, \mn@doi
  [\mnras] {10.1111/j.1745-3933.2012.01287.x}, 424, L64

\bibitem[\protect\citeauthoryear{Mirabal, Charles, Ferrara, Gonthier, Harding,
  Sanchez-Conde  \& Thompson}{Mirabal et~al.}{2016}]{Mirabal_2016}
Mirabal N.,  Charles E.,  Ferrara E.~C.,  Gonthier P.~L.,  Harding A.~K.,
  Sanchez-Conde M.~A.,   Thompson D.~J.,  2016, \mn@doi [\apj]
  {10.3847/0004-637x/825/1/69}, 825, 69

\bibitem[\protect\citeauthoryear{{M{\"o}ller} \& {de
  Boissi{\`e}re}}{{M{\"o}ller} \& {de
  Boissi{\`e}re}}{2020}]{2020MNRAS.491.4277M}
{M{\"o}ller} A.,  {de Boissi{\`e}re} T.,  2020, \mn@doi [\mnras]
  {10.1093/mnras/stz3312}, \href
  {https://ui.adsabs.harvard.edu/abs/2020MNRAS.491.4277M} {491, 4277}

\bibitem[\protect\citeauthoryear{{Naul}, {Bloom}, {P{\'e}rez}  \& {van der
  Walt}}{{Naul} et~al.}{2018}]{2018NatAs...2..151N}
{Naul} B.,  {Bloom} J.~S.,  {P{\'e}rez} F.,   {van der Walt} S.,  2018, \mn@doi
  [Nature Astronomy] {10.1038/s41550-017-0321-z}, \href
  {https://ui.adsabs.harvard.edu/abs/2018NatAs...2..151N} {2, 151}

\bibitem[\protect\citeauthoryear{Padovani et~al.}{Padovani
  et~al.}{2017}]{Padovani:2017zpf}
Padovani P.,  et~al., 2017, \mn@doi [Astron. Astrophys. Rev.]
  {10.1007/s00159-017-0102-9}, 25, 2

\bibitem[\protect\citeauthoryear{{Pasquet}, {Pasquet}, {Chaumont}  \&
  {Fouchez}}{{Pasquet} et~al.}{2019}]{2019A&A...627A..21P}
{Pasquet} J.,  {Pasquet} J.,  {Chaumont} M.,   {Fouchez} D.,  2019, \mn@doi
  [\aap] {10.1051/0004-6361/201834473}, \href
  {https://ui.adsabs.harvard.edu/abs/2019A&A...627A..21P} {627, A21}

\bibitem[\protect\citeauthoryear{Romani}{Romani}{2014}]{Romani159}
Romani R.~W.,  2014, \mn@doi [Science] {10.1126/science.1251943}, 344, 159

\bibitem[\protect\citeauthoryear{Salvetti, Chiaro, La~Mura  \&
  Thompson}{Salvetti et~al.}{2017}]{Salvetti_2017}
Salvetti D.,  Chiaro G.,  La~Mura G.,   Thompson D.~J.,  2017, \mn@doi [\mnras]
  {10.1093/mnras/stx1328}, 470, 1291

\bibitem[\protect\citeauthoryear{Saz~Parkinson, Xu, Yu, Salvetti, Marelli  \&
  Falcone}{Saz~Parkinson et~al.}{2016}]{Saz_Parkinson_2016}
Saz~Parkinson P.~M.,  Xu H.,  Yu P. L.~H.,  Salvetti D.,  Marelli M.,   Falcone
  A.~D.,  2016, \mn@doi [\apj] {10.3847/0004-637x/820/1/8}, 820, 8

\bibitem[\protect\citeauthoryear{{The Fermi-LAT Collaboration}}{{The Fermi-LAT
  Collaboration}}{2019}]{Fermi-LAT:2019pir}
{The Fermi-LAT Collaboration} 2019, {The Fourth Catalog of Active Galactic
  Nuclei Detected by the Fermi Large Area Telescope} (\mn@eprint {arXiv}
  {1905.10771})

\bibitem[\protect\citeauthoryear{{Urry} \& {Padovani}}{{Urry} \&
  {Padovani}}{1995}]{1995PASP..107..803U}
{Urry} C.~M.,  {Padovani} P.,  1995, \mn@doi [PASP] {10.1086/133630}, \href
  {http://adsabs.harvard.edu/abs/1995PASP..107..803U} {107, 803}

\makeatother
\end{thebibliography}

\bsp	
\label{lastpage}
\end{document}